\newcommand\bm{\boldsymbol}

\documentclass[twocolumn,prl,aps,showpacs,floatfix,superscriptaddress,preprintnumbers,amsmath,amssymb]{revtex4}

\usepackage{graphicx}

\begin{document}

\renewcommand{\floatpagefraction}{0.5}

\title{Phase Diagram of the Dzyaloshinskii-Moriya Helimagnet Ba$_2$CuGe$_2$O$_7$ in Canted Magnetic Fields.}

\author{S. M\"uhlbauer \footnote{Present Address: Technische Universit\"at M\"unchen, Forschungsneutronenquelle Heinz Maier-Leibnitz (FRM II), Lichtenbergstr. 1, D-85747 Garching, Germany. Sebastian.muehlbauer@frm2.tum.de}}

\affiliation{Neutron Scattering and Magnetism Group, Laboratory for Solid State Physics, ETH Z\"urich, Z\"urich, Switzerland}

\author{S. Gvasaliya}

\affiliation{Neutron Scattering and Magnetism Group, Laboratory for Solid State Physics, ETH Z\"urich, Z\"urich, Switzerland}

\author{E. Ressouche}

\affiliation{INAC/SPSMS-MDN, CEA/Grenoble, 17 rue des Martyrs, 38054 Grenoble Cedex 9, France}

\author{E. Pomjakushina}

\affiliation{Laboratory for Developments and Methods (LDM), Paul Scherrer Institute, Villingen, Switzerland}

\author{A. Zheludev}

\affiliation{Neutron Scattering and Magnetism Group, Laboratory for Solid State Physics, ETH Z\"urich, Z\"urich, Switzerland}

\date{\today}

\begin{abstract}
The evolution of different magnetic structures of non-centrosymmetric Ba$_2$CuGe$_2$O$_7$ is systematically studied as function of the orientation of the magnetic field $\bm{H}$. Neutron diffraction in combination with measurements of magnetization and specific heat show a virtually identical behaviour of the phase diagram of Ba$_2$CuGe$_2$O$_7$ for $\bm H$ confined in both the (1,0,0) and (1,1,0) plane. The existence of a recently proposed incommensurate double-$k$ AF-cone phase is confirmed in a narrow range for $\bm H$ close to the tetragonal $c$-axis. For large angles enclosed by $\bm H$ and the $c$-axis a complexely distorted non-sinusoidal magnetic structure has recently been observed. We show that its critical field $H_c$ systematically increases for larger canting. Measurements of magnetic susceptibility and specific heat finally indicate the existence of an incommensurate/commensurate transition for ${\bm H} \approx 9\,\,{\rm T}$ applied in the basal $(a,b)$-plane and agree with a non-planar, distorted cycloidal magnetic structure.
\end{abstract}


\vskip2pc

\maketitle

\section{ I. Introduction}

Recently, helical magnetic spin-structures have regained lots of interest, partially motivated by the discovery of topologically stable skyrmion phases in cubic MnSi and other B20 compounds \cite{Muehlbauer:09b,Muenzer:09,Yu:11,Yu:10, Adams:11}. Also the potential multiferroic properties of helimagnets have been recognized in recent years. Theory shows a particularly strong coupling of ferroelectric and ferromagnetic properties in such materials \cite{Mostovoy:06, Sergienko:06}. Spiral magnetic structures are hence speculated to host multiferroicity in numerous different compounds including hexagonal Ba$_{0.5}$Sr$_{1.5}$Zn$_{2}$Fe$_{12}$O$_{22}$ \cite{kimura:05}, perovskite TbMnO$_3$ \cite{Kenzelmann:05} and the kagom\'{e} lattice Ni$_3$V$_2$O$_8$ \cite{Lawes:05}. 

Besides exchange frustration, helical magnetic structures can be promoted by the Dzyaloshinsky-Moriya interaction (DM) \cite{Dzyaloshinskii:58, Moriya:60} which is allowed by symmetry only for antisymmetric exchange paths. Such is the non-centrosymmetric tetragonal antiferromagnet (AF) Ba$_2$CuGe$_2$O$_7$ which shows an incommensurate, almost AF cycloidal spin structure \cite{Zheludev:96}. Surprisingly, unlike the isostructural commensurate weak ferromagnet Ba$_2$CoGe$_2$O$_7$ \cite{Yi:08,Zheludev:03, Sato:03}, Ba$_2$CuGe$_2$O$_7$ shows no signs of spontaneous ferroelectricity \cite{Zheludev:03, Sato:03} in zero field. Moreover, Ba$_2$CuGe$_2$O$_7$ is one of the rare materials where spiral spin textures and weak ferromagnetism are predicted to coexist \cite{Bogdanov:02}. It was suggested by theory that Ba$_2$CuGe$_2$O$_7$ could support stable skyrmion phases \cite{Bogdanov:02} similar to those of the B20 compounds, however such structures have not been observed so far. In contrast, recent neutron diffraction experiments showed the existence of an unexpected AF-cone phase \cite{Muehlbauer:11}. In addition, it turned out that the spin structure of Ba$_2$CuGe$_2$O$_7$ sensitively depends on the orientation of the magnetic field and undergoes several phase transitions \cite{Zheludev:96, Muehlbauer:11}. This paper provides a detailed follow-up study of the recently discovered magnetic structures of Ba$_2$CuGe$_2$O$_7$ \cite{Muehlbauer:11}. In particular, we systematically investigate their evolution as function of the orientation of the magnetic field.

\begin{figure*}
\begin{center}
\includegraphics[width=0.98\textwidth]{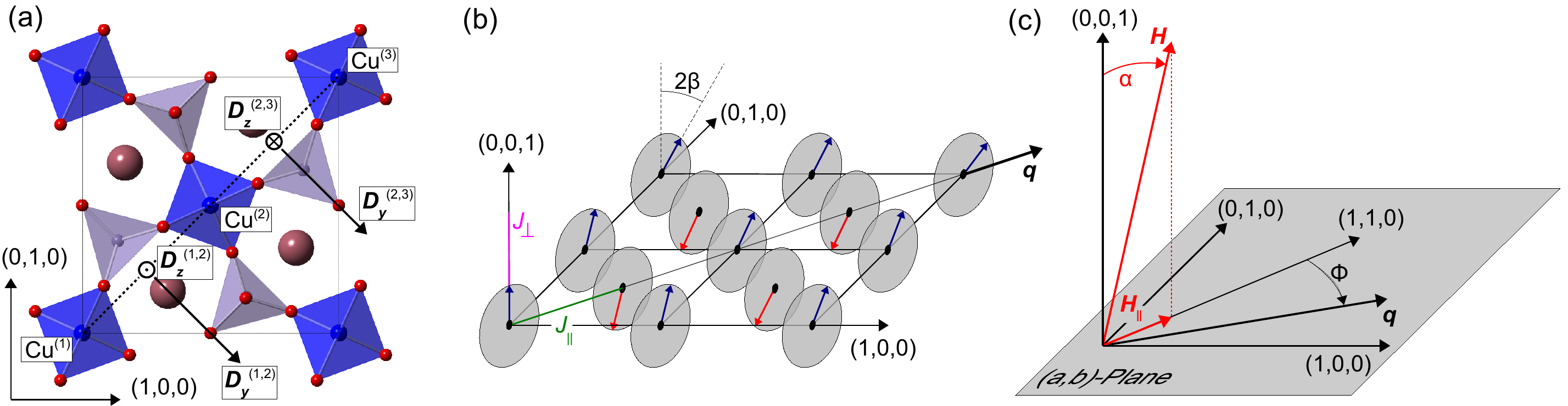}
\caption{Color online Panel (a): Crystallographic structure of Ba$_2$CuGe$_2$O$_7$ viewed along the $c$-axis. The different components of the DM vector are sketched for the Cu-Cu bonds along the (1,1,0) direction (dashed line). The uniform component ${\bm D_y}$ points in the same direction for all bonds. The staggered component ${\bm D_z}$ is sign alternating. Panel (b): Almost AF cycloidal magnetic structure of Ba$_2$CuGe$_2$O$_7$. The dominant exchance paths $J_{\|}$ and $J_{\perp}$ are indicated by the green and magneta colored line, respectively. Panel (c): Sketch of the experimental geometry:  The angle enclosed by the (1,1,0) direction and the propagation vector $\bm q$ is denoted $\phi$, similar to the notation used in Refs. \cite{Zheludev:97b, Muehlbauer:11}. The angle enclosed by the $c$-axis and the magnetic field is denoted $\alpha$. The in-plane component of the magnetic field is denoted $\bm H_\|$.}
\label{Fig_1}
\vspace{-0.03\textwidth}
\end{center}
\end{figure*}

The insulator Ba$_2$CuGe$_2$O$_7$ crystallizes in the non-centrosymmetric tetragonal space group $P{\overline 4}2_1m$ with lattice parameters $a$=8.466\AA, $c$=5.445\AA. A schematic depiction of the unit cell is given in Fig.\,\,\ref{Fig_1}, panel (a). The basic feature of Ba$_2$CuGe$_2$O$_7$ is a square arrangement of Cu$^{2+}$ ions in the $(a,b)$-plane. Nearest-neighbor in-plane AF exchange along the diagonal of the $(a,b)$-plane is the dominant magnetic exchange interaction ($J_{\|}\approx 0.96\,\rm {meV}$ per bond). The interaction between Cu-atoms from adjacent planes is weak and ferromagnetic ($J_{\perp}\approx -0.026\,\rm {meV}$) \cite{Zheludev:99}. The DM vector of Ba$_2$CuGe$_2$O$_7$ consists of two components: For Cu-bonds along the (1,1,0) crystalline direction ${\bm D_y}$ points along the diagonal of the basal $(a,b)$-plane in the (1,-1,0) direction. ${\bm D_z}$ is parallel to the tetragonal $c$-axis and sign-alternating for neighbouring bonds \footnote{Note that the orientation of ${\bm D_y}$ and ${\bm D_z}$ can only be defined in combination with an {\it oriented} Cu-Cu bond as denoted in Fig.\,\,\ref{Fig_1}.}. It has been established by neutron diffraction \cite{ Zheludev:96, Zheludev:97, Zheludev:99, Zheludev:97b,Zheludev:97PB} that the almost AF cycloid, observed below $T_N=3.2\,{\rm K}$ is stabilized by the DM vector ${\bm D_y}$. A schematic depiction of the cycloidal spin-structure is given in Fig.\,\,\ref{Fig_1}, panel (b): For the propagation vector $\bm q$ along (1,1,0), spins are confined in (1,-1,0) plane, the rotation angle $\beta$ relative to a perfect antiparallel alignment is 9.7$^{\circ}$ per unit cell. Due to the equivalency of the (1,$\pm$1,0) directions, two equally populated domains are present. Four magnetic satellite reflections are thus observed at ($1\pm\xi$,$\pm\xi$,0), $\xi\sim0.0273$, centered at the N\'{e}el point (1,0,0). Weak higher order reflections at ($1\pm3\xi$,$\pm3\xi$,0) yield a distortion of the cycloid and can be quantitatively explained taking the KSEA-interaction into account \cite{Zheludev:98bla, Zheludev:99}. 

Cooling below $T_N$ in a small magnetic field of a few tens of Gauss in the $(a,b)$-plane can assure a single domain sample. Note, that for a single domain sample, a small ferroelectric polarisation is indeed observed \cite{Murakawa:09}. Sizable magnetic fields in the $(a,b)$-plane tend to rotate the propagation vector as well as the plane of spin rotation in a rather complicated way, quantitatively explained by the interplay of tetragonal crystal anisotropy and Zeemann energy \cite{Zheludev:97b}. For magnetic field $\bm H$ along the  $c$-axis, the almost cycloidal spin structure of Ba$_2$CuGe$_2$O$_7$ distorts to a soliton lattice with divergent periodicity until an incommensurate/commensurate (I/C) transition is observed at $\approx 2.4\,{\rm T}$ \cite{Zheludev:97, Zheludev:99}. However, we could already show that this model of Ba$_2$CuGe$_2$O$_7$ is still fairly incomplete \cite{Muehlbauer:11}: Close to the I/C transition, an intermediate phase of unknown origin has been attributed to a distinct phase with a peculiar AF-cone 2$k$-structure \cite{Muehlbauer:11}. The AF-cone phase was found for a narrow range of magnetic fields aligned closely to the crystallographic $c$-axis. In contrast, for a large misalignment of $\bm H$, a crossover to a complexely distorted non-sinusoidal phase was speculated \cite{Muehlbauer:11}.

We have systematically examined the evolution of these recently discovered magnetic phases of Ba$_2$CuGe$_2$O$_7$ for magnetic fields in both the (1,1,0) and (1,0,0) plane. We use neutron diffraction in combination with bulk measurements of magnetiztion and specific heat. Our measurements indicate a virtually identical behaviour for magnetic field confined in the (1,0,0) and (1,1,0) plane. We confirm the existence of the AF-cone phase in a narrow range of $\lesssim \pm 10^{\circ}$ of $\bm H$ with respect to the $c$-axis. We further show that the critical field $H_c$ of the distorted phase systematically increases for larger canting angles. Measurements of magnetization and specific heat finally indicate the existence of a unexpected I/C transition for magnetic fields applied in the basal $(a,b)$-plane at $H_c\approx 9\,\,{\rm T}$. The observation of this transition in combination with both odd and even harmonics of $\bm q$ at $2\xi$ and $3\xi$ agrees with a non-planar, distorted magnetic structure, most likely caused by the staggered component ${\bm D_z}$ of the DM vector.

\section{II. Experimental Methods}

Two single crystal samples with a length of $\sim 50\,{\rm mm}$ and a diameter of $\sim 5\,{\rm mm}$ (denoted {\it A} and {\it B}) of Ba$_2$CuGe$_2$O$_7$ were grown with a floating zone image furnace. The starting material has been obtained by a standard solid state reaction at 880-1000$^{\circ}$C using starting materials of Ba(CO$_3$)$_2$, CuO and GeO$_2$ with a purity of 99.99\%. The phase purity of the compound was checked with a conventional X-ray powder diffractometer. The resulting powder was hydrostatically pressed in the form of rods (8 mm in diameter and  $\approx\,80\,{\rm mm}$ in length) and sintered at 980$^{\circ}$C. The crystal has been annealed after the growth to release possible internal stress. The bulk material of both samples exhibits a transparent yellowish color, however, sample {\it A} showed a slightly darker surface.

For the measurements of the magnetization, a small cube with (1,0,0) facets and a volume of 1\,mm$^3$ has been cut from crystal {\it A} with a diamond wire saw. A similar cube with (1,1,0) facets has been cut from crystal {\it B}. The measurements of susceptibility have been performed with a SQUID magnetometer (Quantum Design MPMS XL 7T) for magnetic fields below 7\,T and a vibrating sample magnetometer (Quantum Design PPMS 14T) for fields above 7\,T. The magnetic field $\bm H$ was applied both in a (1,0,0) and a (1,1,0) plane, inclined at the angle $\alpha$ with respect to the $c$-axis. $\alpha$ was varied from 0$^{\circ}$ to 90$^{\circ}$ with a precision of $\pm 2^{\circ}$. 

The specific heat was measured with the relaxation method using a Quantum Design PPMS 14T. Three thin platelets with a size of 1x1x0.2\,mm$^3$ have been cut from crystal {\it B}. The flat geometry of the samples has been chosen to ensure optimal thermal coupling between the sample and the puck platform. The samples have been mounted to the puck platform with a small amout of Apiezon N grease. The contribution of the grease to the specific heat has been determined separately and was subtracted accordingly. The magnetic field was applied parallel to the normal vector of the platelets which was oriented parallel to the $c$-axis for the first sample. The normal vector of the second and third platelet were inclided with respect to the $c$-axis at an angle of $\alpha=25^{\circ}$ and $\alpha=90^{\circ}$ in a (1,1,0) plane, respectively.

For the neutron diffraction experiments, a large part of crystal {\it B} with a length of  $\sim 15\,{\rm mm}$ has been used. Neutron diffraction experiments have been performed on the triple axis spectrometer TASP at PSI \cite{semadeni:01} and on the lifting counter diffractometer D23 at ILL. A split-coil cryomagnet covering a range of $T=1.6\,{\rm K}$ to $3.5\,{\rm K}$ and $H=0\,{\rm T}$ to $6\,{\rm T}$ was used on both instruments. TASP was configured in elastic mode with high $Q$ resolution at an incident wavevector of $1.3\,$\AA$^{-1}$. An {\it open-20'-20'-20'} configuration was used for most of the scans except the scans covering the higher order diffraction peaks at $3\xi$, where {\it open-80'-80'-80'} was used to relax the instrumental resolution for the sake of intensity. On D23, an incident wavevector of $2.65\,$\AA$^{-1}$ from a graphite monochromator has been used. The resolution of D23 was defined by a Cadmium hole mask of $\sim 6\,{\rm mm}$ in diameter before the sample and a Cadmium hole mask of $\sim 10\,{\rm mm}$ before the detector.

A non-magnetic micro-goniometer inside the cryomagnet was used to mount the sample, allowing to tilt the $c$-axis with respect to the magnetic field. The sample was pre-aligned using X-ray Laue diffraction with a precision of 2$^{\circ}$ prior to the experiment. The precise alignment of the sample was confirmed by in-situ neutron diffraction. The nomenclature used to describe the geometry of our experimental setup is given in Fig.\,\,\ref{Fig_1}, panel (c), similar to the notation used in References \cite{Zheludev:97b, Muehlbauer:11}: The angle enclosed by the crystallographic (1,1,0) direction and the propagation vector $\bm q$ is denoted $\phi$. The angle enclosed by the magnetic field $\bm H$ and the $c$-axis is denoted $\alpha$. The in-plane component of the magnetic field is denoted $\bm H_\|$.

The magnetic field was applied both in a (1,0,0) and a (1,1,0) plane. Limited by the maximal tilt of the cryomagnet and the microgoniometer, neutron diffraction data was collected on TASP for $\alpha=5^{\circ}$, confined in both the (1,0,0) and the (1,1,0) plane and for $\alpha=15^{\circ}$ confined in the (1,0,0) plane. On D23, data was collected for $\alpha=15^{\circ}$ and $\alpha=30^{\circ}$, confined in the (1,1,0) plane. For each magnetic field neutron diffraction data was collected after field-cooling from $T=6\,{\rm K}$ to the base temperature of $T=1.65\,{\rm K}$. Elastic scans have been performed along the line in reciprocal space given by $(1+\sqrt{2}\zeta \cos (\phi-\pi/4),-\sqrt{2}\zeta \sin (\phi-\pi/4),0)$, $\zeta$ being the scan parameter \footnote{The expression given in Ref.\,\,\cite{Muehlbauer:11} defining the scans in reciprocal space is lacking the prefactor of $\sqrt{2}$.}.

\section{III. Experimental Results}

\subsection{Magnetic susceptibility}

\begin{figure*}
\begin{center}
\includegraphics[width=0.92\textwidth]{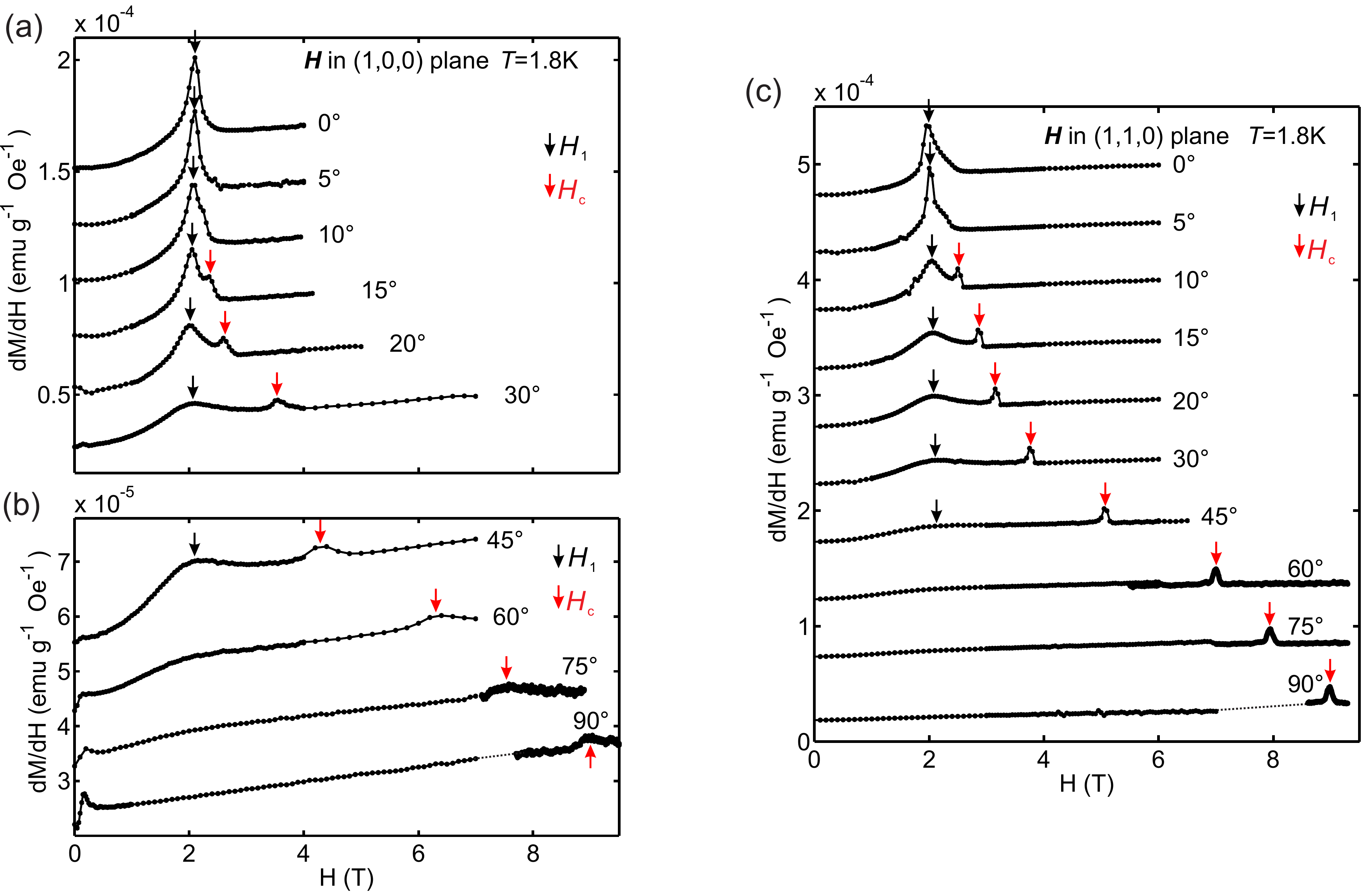}
\caption{Color online Panel (a): Magnetic field dependence of the static magnetic susceptibility for canted magnetic fields in the (1,0,0) plane for $\alpha$ from $0^{\circ}$ to 30$^{\circ}$. The data was taken at $T=1.8\,{\rm K}$. The curves have been shifted vertically by $0.25*10^{-4}\,{\rm emu\,g^{-1} Oe^{-1}}$ for clarity. Panel (b): Similar data obtained for $\alpha$ from $45^{\circ}$ to 90$^{\circ}$, shifted  by $0.1*10^{-4}\,{\rm emu\,g^{-1} Oe^{-1}}$. Panel (c): Field dependence of the static magnetic susceptibility for canted magnetic fields in the (1,1,0) plane. The curves have been shifted  by $0.5*10^{-4}\,{\rm emu\,g^{-1} Oe^{-1}}$.}
\label{Fig_2}
\vspace{-0.03\textwidth}
\end{center}
\end{figure*}

We start with a description of $\rm d \bm M/\rm d \bm H$, given in Fig.\,\,\ref{Fig_2}. The data was taken at a temperature of $T=1.8\,{\rm K}$. Panels (a) and (b) show the magnetic field dependence for $\alpha$ from $0^{\circ}$ to 90$^{\circ}$, applied in the (1,0,0) plane. For magnetic field $\bm H$ parallel to the $c$-axis ($\alpha=0^{\circ}$), a single sharp peak is seen at $H_{1}=2.05\,{\rm T}$. In contradiction with the interpretation put forward in Ref.\,\, \cite{Zheludev:97}, the sharp anomaly does {\it not} correspond to the I/C transition but indicates the transition from the soliton lattice to the newly discovered AF-cone phase seen in neutron diffraction at $T=1.65\,{\rm K}$ and $H_{1}=1.95\,{\rm T}$ \cite{Muehlbauer:11}. The I/C transition itself at $H_{c}\sim2.4\,{\rm T}$ is featureless in $\rm d \bm M/\rm d \bm H$. Virtually identical behaviour is observed for $\alpha=5^{\circ}$ and $\alpha=10^{\circ}$. 

The behaviour qualitatively changes at $\alpha=15^{\circ}$, where the peak splits into a feature at higher fields $H_{c}$ while the transition seen at $H_{1}$ simultaneously broadens. The sharp feature seen at $H_{c}=2.4\,{\rm T}$ is identified as the I/C transition seen in neutron diffraction for $\alpha=15^{\circ}$ \cite{Muehlbauer:11}. In susceptibility, it can be followed until it reaches $H_{c}=9\,{\rm T}$ for $\alpha=90^{\circ}$. The broadened feature at $H_{1}$ corresponds to the crossover from the soliton lattice to the proposed distorted structure seen in neutron diffraction \cite{Muehlbauer:11}. For $\alpha\gtrsim45^{\circ}$ the broadened anomaly at $H_1$ is completely smeared out in magnetic susceptibility.

The field dependence of  $\rm d \bm M/\rm d \bm H$ for $\alpha$ from $0^{\circ}$ to 90$^{\circ}$, applied in the (1,1,0) plane is shown in panel (c) of Fig.\,\,\ref{Fig_2}. The same qualitative behaviour is observed for magnetic field confined in the (1,1,0) plane as compared to the (1,0,0) plane. However, data obtained on the sample cut from the crystal {\it B} shows features in more clarity, in particular for large $\alpha$. This behaviour is attributed to a better crystallograhic quality of sample {\it B}. The slight deviations seen for the transition fields $H_1$ and $H_c$ are attributed to the finite precision of the sample alignment.

\subsection{Specific heat}

\begin{figure}
\begin{center}
\includegraphics[width=0.45\textwidth]{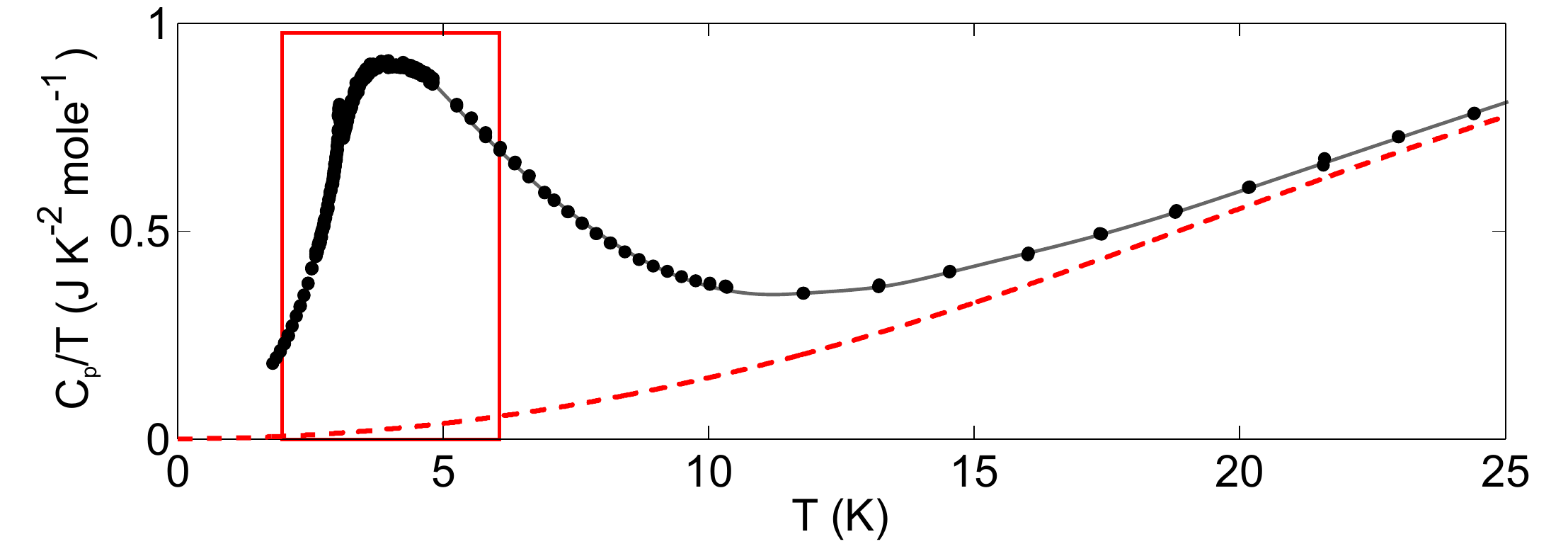}
\caption{Color online Specific heat $C_p/T$ of Ba$_2$CuGe$_2$O$_7$ for $H=0\,{\rm T}$. The red box corresponds to the data shown in detail in Fig.\,\,\ref{Fig_3_b}. The dashed red line corresponds to the phonon contribution obtained by a fit to the Debye model.}
\label{Fig_3_a}
\vspace{-0.02\textwidth}
\end{center}
\end{figure}

The 2$^{\rm nd}$ order phase transition from the paramagnetic phase to the 3D long-range-ordered (LRO) spiral phase at $T_N=3.2\,{\rm K}$ in $H=0\,{\rm T}$ leads to a signature in specific heat. Fig.\,\,\ref{Fig_3_a} shows the specific heat $C_p/T$ of a small platelet of Ba$_2$CuGe$_2$O$_7$ for $H=0\,{\rm T}$ from $T=2\,{\rm K}$ to $T=25\,{\rm K}$.

Below $T=10\,{\rm K}$ a broad maximum of the specific heat, centered at $T=4.5\,{\rm K}$, is visible. At $T_N=3.2\,{\rm K}$ a small lambda anomaly, seen on top of the broad peak corresponds to the spiral LRO. The phonon contribution has been fitted to a Debye model between 10\,K and 50\,K, yielding $\Theta_D=177\,\,{\rm K}$ and $n_{effective}=4.18$. The phonon contribution is indicated by the dashed red line in Fig.\,\,\ref{Fig_3_a}, clearly indicating a significant magnetic contribution. The phonon part was subtracted accordingly.

\begin{figure*}
\begin{center}
\includegraphics[width=0.99\textwidth]{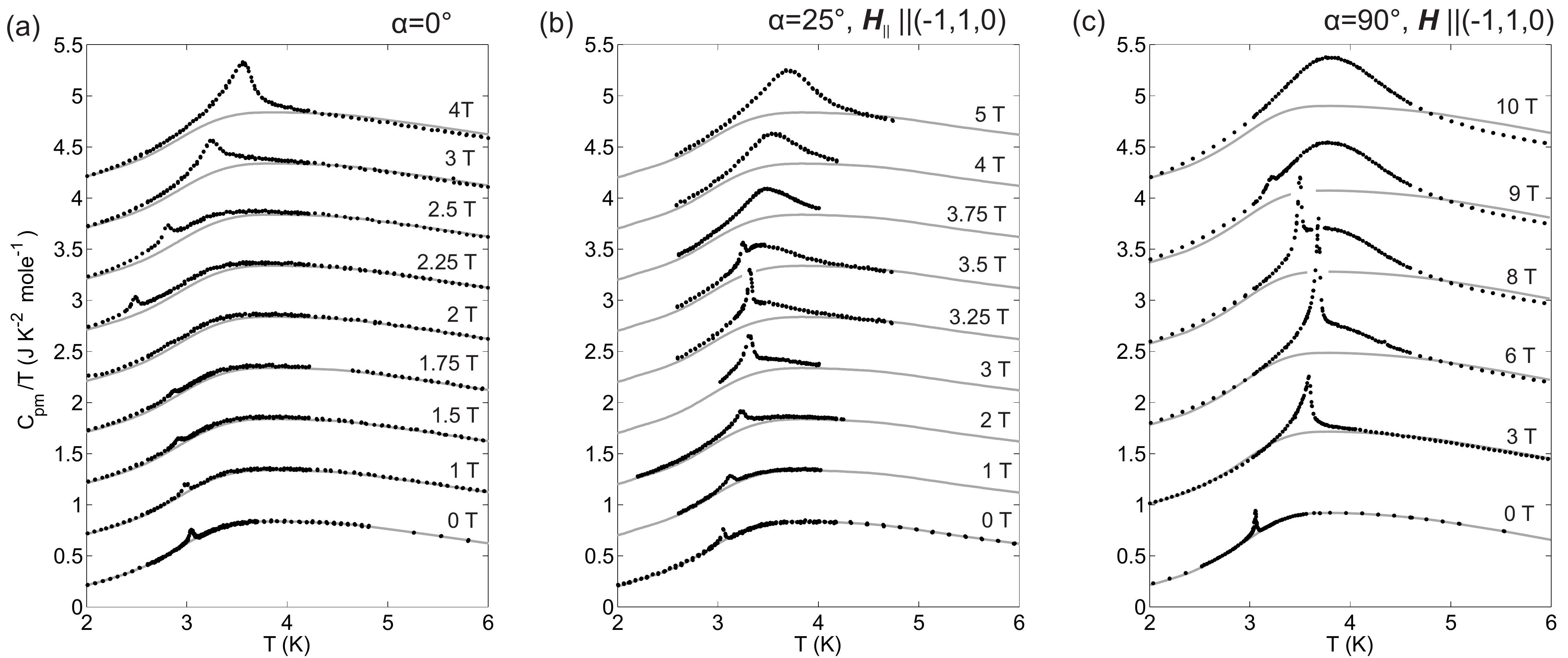}
\caption{Color online Temperature dependence of the magnetic specific heat $C_{pm}/T$ of Ba$_2$CuGe$_2$O$_7$ for different magnetic fields. The grey lines correspond to the data obtained for zero field and serve as guide to the eye. Panel (a): Magnetic field along the $c$-axis ($\alpha=0^{\circ}$). Panel (b): Data obtained for $\alpha=25^{\circ}$ confined in the (1,1,0) plane. The curves in panel (a) and panel (b) have been shifted by $0.5\,{\rm J K^{-2} mole^{-1}}$ for clarity. Panel (c): Magnetic field applied in the (-1,1,0) direction ($\alpha=90^{\circ})$. The curves have been shifted by $0.8\,{\rm J K^{-2} mole^{-1}}$ for clarity. }
\vspace{-0.03\textwidth}
\label{Fig_3_b}
\end{center}
\end{figure*}

The remaining magnetic contribution to the specific heat $C_{pm}/T$ is shown in greater detail in Fig.\,\,\ref{Fig_3_b}, panel (a) for temperatures between 2\,K and 6\,K and different magnetic fields $\bm H$ aligned along the $c$-axis ($\alpha=0^{\circ}$). As already introduced in the previous paragraph, the specific heat is governed by a small anomaly at $T_N=3.2\,{\rm K}$ on top of a broad background maximum, centered at $T\sim4.5\,{\rm K}$. For increasing magnetic field $H<H_c(T)$ the lambda anomaly at $T_N$ slightly shifts to lower temperatures. Simultaneously, the weight of the anomaly decreases. At the critical field $H_c=2\,{\rm T}$, no signs of a lambda anomaly could be observed any more. However, for magnetic fields above the critical field $H>H_c(T)$ a significantly broadened peak is observed at $T_N$ instead of the lambda anomaly. Both the weight and the transition temperature of the broadened peak increase for further increasing magnetic field. The broad background maximum is only marginally affected by the magnetic field along the $c$-axis.

Panel (b) of Fig.\,\,\ref{Fig_3_b} shows similar data obtained on a sample where the magnetic field $\bm H$ was tilted by $\alpha=25^{\circ}$ in a (1,1,0) plane. In contrast to the setup with $\alpha=0^{\circ}$, an increased weight of the lambda anomaly is observed with increasing magnetic field. Furthermore, an upward shift of the transition temperature is associated with increasing field until $H=3.25\,{\rm T}$. At the critical field $H_c\approx3.5\,{\rm T}$ the lambda anomaly significantly shrinks and shifts to lower temperatures while a broader maximum of $C_{pm}/T$ appears at slightly higher temperature. For magnetic fields $H>3.5\,{\rm T}$ no lambda anomaly is observed any more. Instead, the weight of the broadened peak again increases and its maximum´s temperature simultaneously shifts upward. Due to the limited scan range no conclusion can be drawn how the broad background maximum is affected by the tilted magnetic field.

The characteristic behaviour of $C_{pm}/T$ for magnetic field $\bm H$ aligned along the (1,-1,0) axis ($\alpha=90^{\circ}$) is given in panel (c) of Fig.\,\,\ref{Fig_3_b}. Similar to the trend observed for $\alpha=25^{\circ}$ both weight and transition temperature of the lambda anomaly further increase with increasing magnetic field until they collapse at the critical field $H_c\approx9\,{\rm T}$, consistent with susceptibility. For magnetic fields above $H>9\,{\rm T}$ no lambda anomaly is observed any more. Again, the lambda-like transition is replaced by a broadened peak which continuously emerges above $H=6\,{\rm T}$. In contrast to panel (a), where the broad background maximum is only marginally affected by the magnetic field, a significantly reduced magnetic specific heat $C_{pm}/T$ is observed for temperatures above $T_N$ for high magnetic fields aligned along the (1,-1,0) axis.

\subsection{Neutron diffraction.}

First neutron diffraction experiments for magnetic field $\bm H$ strictly parallel to the $c$-axis ($\alpha=0^{\circ}$) and for $\bm H$ confined in the (1,0,0) \footnote{Correction: After publication of Ref.\,\,\cite{Muehlbauer:11} we became aware of an inconsistency in our notation: The data presented in \cite{Muehlbauer:11} was obtained with the magnetic field confined in the (1,0,0) plane and not the (0,1,0) plane.} plane have already been discussed Ref.\,\,\cite{Muehlbauer:11}. However, for $\alpha=5^{\circ}$ and $\alpha=15^{\circ}$, data was shown only for one of the twin domains . Here, we extend the data presented in Ref.\,\,\cite{Muehlbauer:11} to the second domain in this geometry. Moreover, diffraction data recorded for magnetic field confined in the (1,1,0) plane allows for a direct comparison between the different geometries.

\subsubsection{Neutron diffraction for small $\alpha$.}

We first discuss our results for a small tilt of magnetic field $\bm H$ by $\alpha=5^{\circ}$ in both the (1,0,0) and (1,1,0) plane. For  $\alpha=0^{\circ}$ \cite{Muehlbauer:11} the magnetic field depedence was governed by (i) the gradual distortion of the cycloidal spin-structure to the solition lattice \cite{Zheludev:99}, leading to an increased weight of the higher order satellite reflections at $\pm 3\xi$ and (ii) by the phase transition to the recently observed incommensurate AF-cone phase at $H_1=1.95\,{\rm T}$ \cite{Muehlbauer:11}. Characteristic of the phase transition to the AF-cone phase is  the rotation of the propagation vector $\bm q$ by $\pi/2$, the appearance of the commensurate (1,0,0) reflection and the missing higher order reflections at $3\xi$. The I/C transition takes place at a higher field of $H_c\approx 2.4\,{\rm T}$.

\begin{figure*}
\begin{center}
\includegraphics[width=0.97\textwidth]{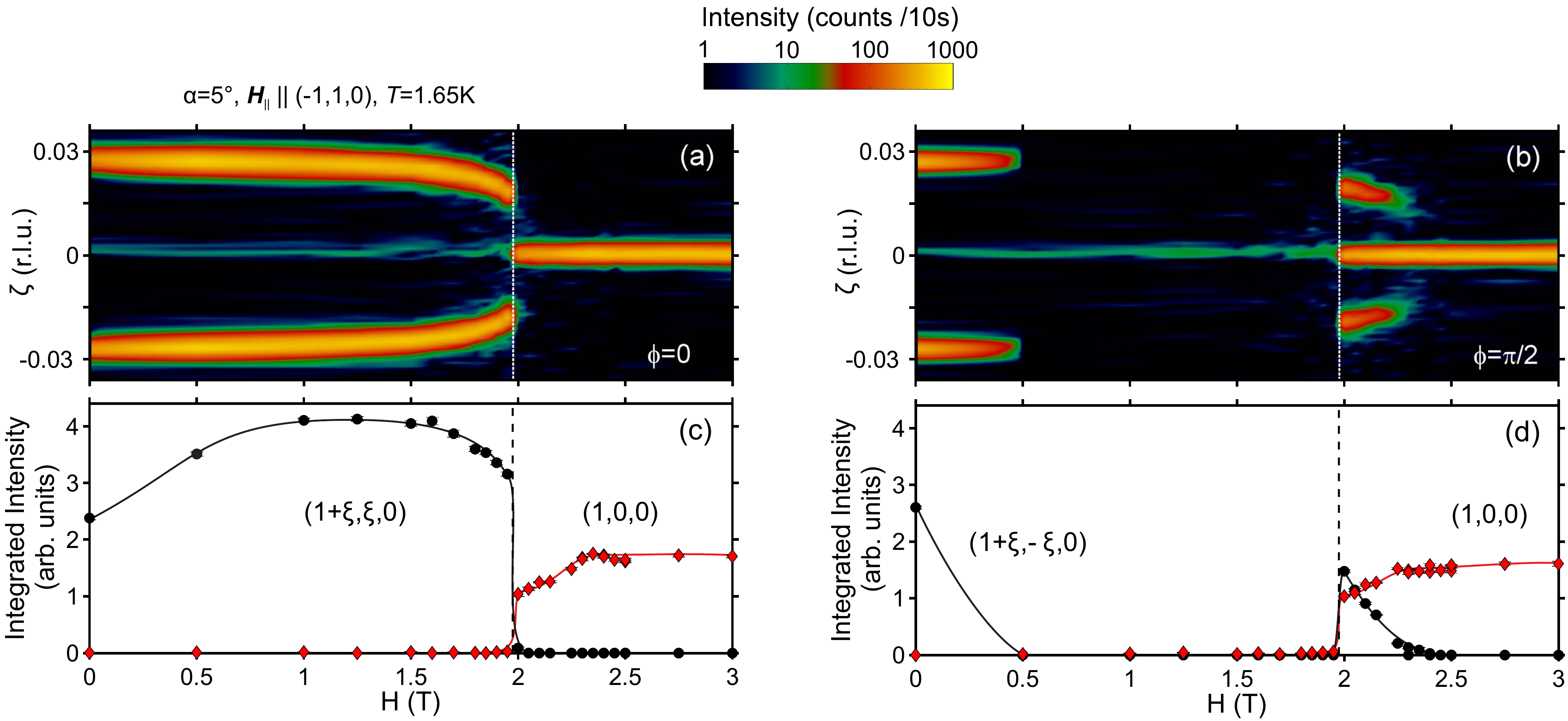}
\caption{Color online Neutron diffraction data as function of magnetic field for $\alpha=5^{\circ}$ and $T=1.65\,{\rm K}$. The magnetic field is confined in the (1,1,0) plane corresponding to $\bm H_\|$ along (-1,1,0). Magnetic field dependence of the incommensurate satellite peaks around (1,0,0) for $\phi=0^{\circ}$ (panel (a)) and  $\phi=90^{\circ}$ (panel (b)). Field dependence of the integrated intensity of the incommensurate satellite reflections at $\xi$ and the commensurate magnetic Bragg peak at (1,0,0) for $\phi=0^{\circ}$ (panel (c)) and  $\phi=90^{\circ}$ (panel (d)), as obtained from Gaussian fits. The lines serve as guide to the eye.}
\label{Fig_5}
\vspace{-0.03\textwidth}
\end{center}
\end{figure*} 

\begin{figure*}
\begin{center}
\includegraphics[width=0.97\textwidth]{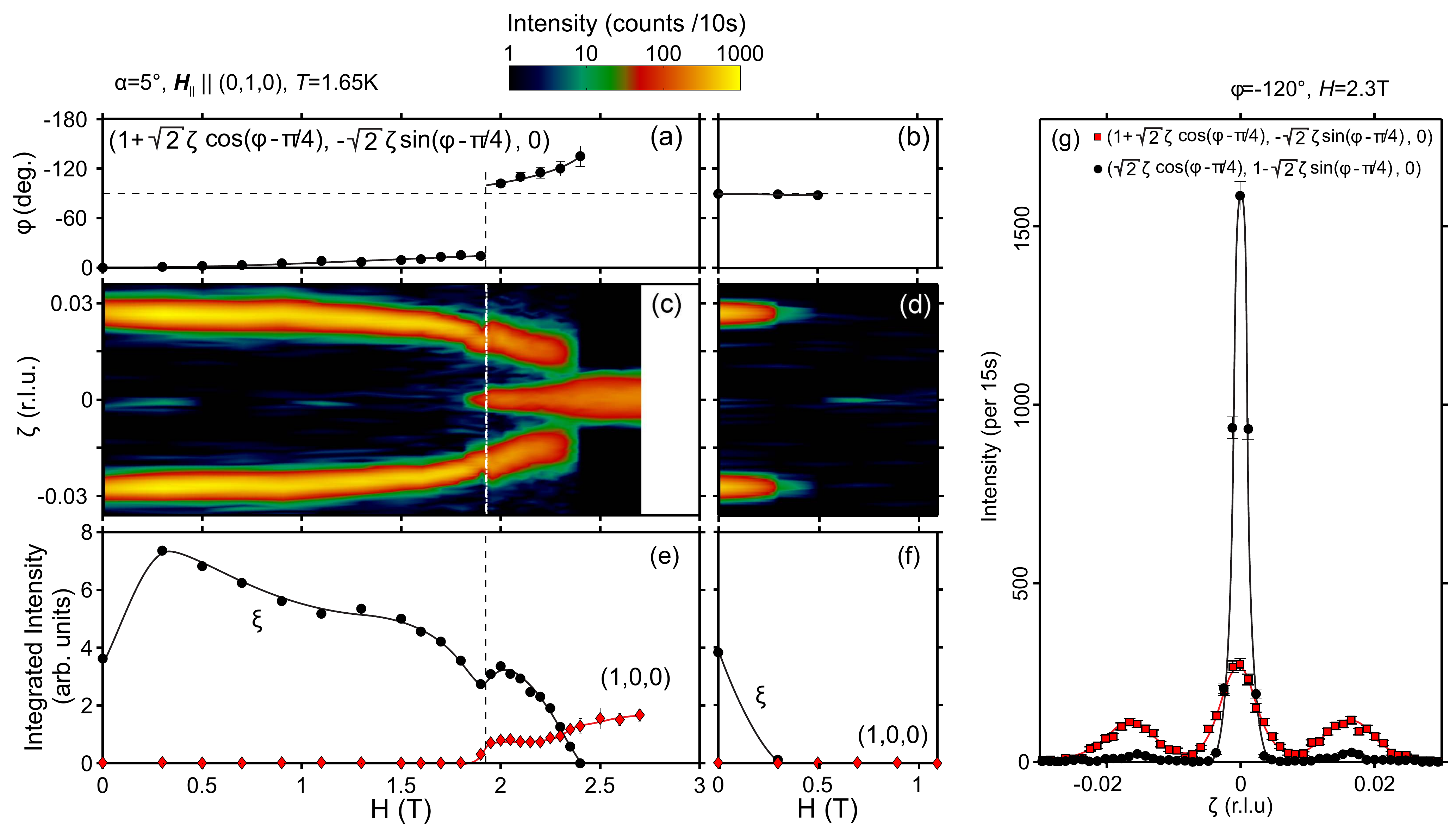}
\caption{Color online Neutron diffraction data as function of magnetic field for  $\alpha=5^{\circ}$ and $T=1.65\,{\rm K}$. The magnetic field is confined in the (1,0,0) plane corresponding to $\bm H_\|$ along (0,1,0). Panel (a) and panel (b) show the magnetic field dependence of $\phi$ for both domains. Panel (c) and panel (d) depict neutron intensity measured in linear scans along $(1+\zeta \sqrt{2}\cos (\phi-\pi/4), -\zeta \sqrt{2} \sin (\phi-\pi/4),0)$ with $\phi$ given in panels (a) and (b), respectively. Panels (e) and (f): Field dependence of the integrated intensity of the incommensurate satellite reflections at $\xi$ and the commensurate magnetic Bragg peak at (1,0,0), as obtained from Gaussian fits. Panel (g) shows scans crossing the incommensurate satellite reflections centered at the  N\'{e}el point (1,0,0) in red and (0,1,0) in black, respectively. The data were taken at $H=2.3\,{\rm T}$, corresponding to $\phi=-120^{\circ}$. The data in (a), (c) and (e) are taken from Ref.\,\,\cite{Muehlbauer:11} [34-36]. The lines serve as guide to the eye.}
\label{Fig_6}
\vspace{-0.03\textwidth}
\end{center}
\end{figure*}

The neutron diffraction data obtained for a misalignment of the magnetic field in the (1,1,0) plane by $\alpha=5^{\circ}$ is given in Fig.\,\,\ref{Fig_5}: The magnetic field dependence of the incommensurate satellite peaks at ($1\pm\xi$,$\pm\xi$,0) for $\phi=0^{\circ}$ and $\phi=90^{\circ}$ is given in the color maps in panels (a) and (b), respectively. The false-color plots are constructed of scans performed along $(1\pm\zeta,\pm\zeta,0)$ for various fields as given in panels (c) and (d). The color maps have been obtained after an interpolation of the diffraction data. The integrated intensity of both the incommensurate peaks and the commensurate (1,0,0) reflection is given in panels (c) and (d). Already for small magnetic fields $H\approx0.5\,{\rm T}$ the domain with its propagation vector parallel to the in-plane component of the magnetic field ($\bm H_\|$ along (-1,1,0)) is depopulated ($\phi=90^{\circ}$). For magnetic fields $H>0.5\,{\rm T}$ the sample is hence in a single domain state with its plane of spin rotation perpendicular to $\bm H_\|$. The phase transition from the soliton lattice to the AF-cone phase at $H_1=1.95\,{\rm T}$ can clearly been identified, similar to the setup with $\alpha=0^{\circ}$ \cite{Muehlbauer:11}: The propagation vector $\bm q$ flips by $\pi/2$ from $\phi=0^{\circ}$ to $\phi=90^{\circ}$ and the commensurate reflection at (1,0,0) rapidly gains intensity. The I/C transition is seen at $H_c\approx 2.4\,{\rm T}$.

Diffraction data of one of the two domains for a misalignment of $\alpha=5^{\circ}$ and magnetic field applied in the (1,0,0) plane [35] has already been discussed in Ref.\,\,\cite{Muehlbauer:11}. For completeless, we reproduce the same data \footnote{Due to a trivial mistake, the error bars shown in the corresponding plots in Ref.\,\,\cite{Muehlbauer:11} are considerably over-estimated. The error bars shown in the present work have been corrected.} in panels (a), (c) and (e) of Fig.\,\,\ref{Fig_6}, allowing for a direct comparison with measurements on the 2$^{nd}$ domain in this geometry given in  panels (b), (d) and (f). Due to the in-plane field component $\bm H_\|$ along (0,1,0), a gradual rotation of the propagation vector $\bm q$ is observed \cite{Zheludev:97b}. The direction of the propagation vector was determined by centering the incommensurate reflections for each field. Panels (a) and (b) show the resulting magnetic field dependence of $\phi$ for both domains. The scans that make up the false-color plot in panels (c) and (d) are thus performed along $(1+\sqrt{2}\zeta \cos(\phi-\pi/4), -\sqrt{2}\zeta \sin(\phi-\pi/4),0)$ [34] with a field-dependent $\phi$ as given in panels (a) and (b).

It is found that the in-plane component $\bm H_\|$ again is responsible for a single domain sample above $\approx 0.5\,{\rm T}$. At the same time, a gradual rotation of the incommensurate satellite peak around the $(1,0,0)$ position towards the direction of $\bm H_\|$ is observed in agreement with Ref.\,\,\cite{Zheludev:97b}, changing from $\phi=0^{\circ}$ at $H=0\,{\rm T}$ to $\phi\approx -15^\circ$ at $H=1.9\,{\rm T}$. At $H_1=1.93\,{\rm T}$, a flip of the propagation vector by precisely -90 degrees ($\phi\rightarrow \phi-\pi/2$ ) is observed on top of the gradual rotation, characteristic of the phase transition to the AF-cone phase \cite{Muehlbauer:11}. The magnetic field dependence of the integrated intensity of the commensurate reflection at (1,0,0) and the incommensurate satellite peaks is depicted in panels (e) and (f) of Fig.\,\,\ref{Fig_6}. The intensity of the commensurate reflection increases in the form of two steps, similar to $\alpha=0^{\circ}$ \cite{Muehlbauer:11}. The slight jump seen in the integrated intensity of the satellite reflections, however, 
can be attributed to polarization effects caused by the different position in reciprocal space or the different magnetic structure of the soliton lattice and the AF-cone phase. Again, the I/C transition is finally observed at a magnetic field of $H_c\approx 2.4\,{\rm T}$.

A closer insight into the magnetic structure of the AF-cone phase can be drawn from this configuration. The field-dependent direction of the propagation vector $\bm q$ is $\phi\neq 45^{\circ}$, allowing to exploit the polarization factor of a neutron diffraction experiment. Neutrons are only scattered by components of the magnetization perpendicular to the momentum transfer ${\bm Q}$. Measurements of the AF-cone phase centered at both the (1,0,0) and (0,1,0) N\'{e}el point indicate that the incommensurate component of the 2$k$-structure is oriented perpendicular to its commensurate AF component. The commensurate AF component is thereby aligned perpendicular to the in-plane component of the magnetic field $\bm H_\|$ along (0,1,0). Typical data plots are shown in  panel (g) for $H=2.3\,{\rm T}$, corresponding to $\phi=-120^{\circ}$. Red markers indentify scans centered at the (1,0,0) N\'{e}el point, black markers indicate scans centered at (0,1,0), respectively.

\subsubsection{Neutron diffraction for large $\alpha$.}

\begin{figure*}
\begin{center}
\includegraphics[width=0.97\textwidth]{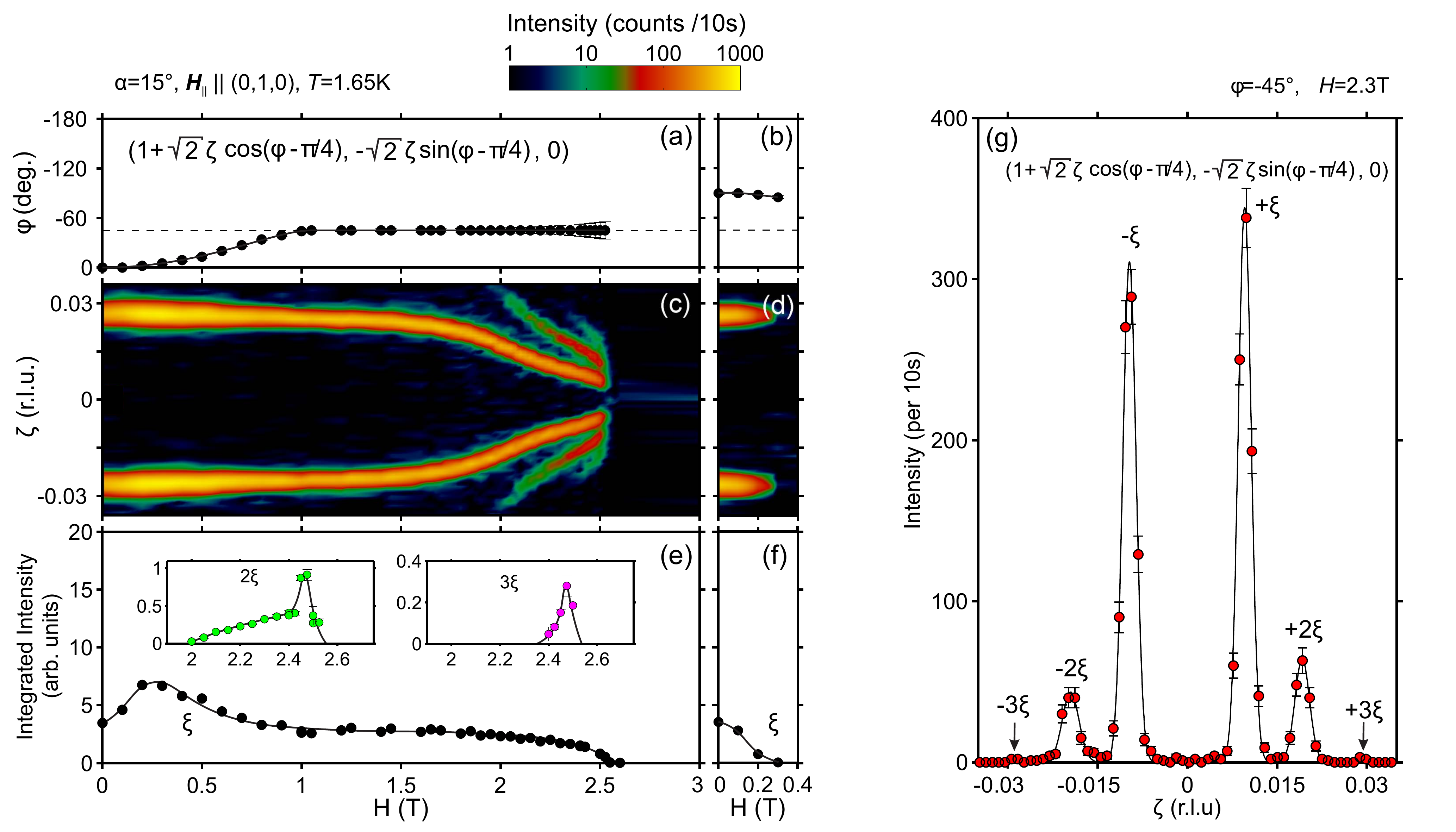}
\caption{Color online Neutron diffraction data as function of magnetic field for  $\alpha=15^{\circ}$ and $T=1.65\,{\rm K}$. The magnetic field is confined in the (1,0,0) plane corresponding to $\bm H_\|$ along (0,1,0). Panels (a) and (b) show the magnetic field dependence of $\phi$ for both domains. Panels (c) and (d) depict neutron diffraction intensity measured in linear scans along $(1+\zeta \sqrt{2} \cos (\phi-\pi/4), -\zeta \sqrt{2} \sin (\phi-\pi/4),0)$ with $\phi$ given in panels (a) and (b), respectively. Panels (e) and (f): Field dependence of the integrated intensity of the incommensurate satellite reflections at $\xi$ and the higher order reflections observed at $2\xi$ and $3\xi$, as obtained from Gaussian fits. A typical scan at $H=2.3\,{\rm T}$, showing even and odd multilies of $\bm q$ at  $2\xi$ and $3\xi$ is depicted in panel (g). The data in (a), (c) and (e) are taken from Ref.\,\,\cite{Muehlbauer:11} [34-36]. The lines serve as guide to the eye.}
\label{Fig_7}
\vspace{-0.02\textwidth}
\end{center}
\end{figure*}

The results of the neutron diffraction experiments for a large tilt of $\bm H$ with respect to the $c$-axis are discussed in the following paragraphs. 
Again, the behavior for one of the domains in the case of $\alpha=15^{\circ}$ and magnetic field in the (1,0,0) plane [35](shown in panels (a), (c) and (e) of Fig.\,\,\ref{Fig_7} [36]) has already been discussed in Ref.\,\,\cite{Muehlbauer:11}, but is shown here anew for a direct comparison with the data for its twin domain (panels (b), (d) and (f)) and for completeness. Similar to the previous setup ($\alpha=5^{\circ}$), a gradual rotation of the propagation vector $\bm q$ is caused by the in-plane field component $\bm H_\|$ along (0,1,0). Again, the scans that make up the false-color plot in panels (c) and (d) are  performed along $(1+\sqrt{2}\zeta \cos(\phi-\pi/4), -\sqrt{2}\zeta \sin(\phi-\pi/4),0)$ [34] with a field-dependent $\phi$ as given in panels (a) and (b). Panels (e) and (f) show the magnetic field dependence of the integrated intensity of the satellite reflection at $\xi$ and the higher-order diffraction peaks at $2\xi$ and $3\xi$, as obtained from Gaussian fits. Note, that the absence of the commensurate (1,0,0) reflection is due to the polarization factor: As $\bm H_\|$ is oriented along (0,1,0), commensurate AF spins are aligned along the (1,0,0) direction and thus do not contribute to magnetic scattering.

For magnetic fields $H\geq 0.3\,{\rm T}$ a single domain state is reached. For magnetic fields $H\geq 1\,{\rm T}$ the propagation vector $\bm q$ is fully rotated into the direction of $\bm H_\|$. In contrast to the data obtained for zero or small $\alpha$, no abrupt reorientation of the propagation vector is observed for $\alpha=15^{\circ}$. As can be seen from panels (c) and (e) the behavior for $H>2\,{\rm T}$ is characterized by the smooth appearance of higher order reflections at even and odd multiplies at 2$\xi$ and 3$\xi$ (inset of panel (e)). The I/C transition is seen at $H_c=2.6\,{\rm T}$.  A typical scan at $H=2.3\,{\rm T}$, showing even and odd multiplies of $\bm q$ at  $2\xi$ and $3\xi$ is depicted in panel (g).

\begin{figure}
\begin{center}
\includegraphics[width=0.48\textwidth]{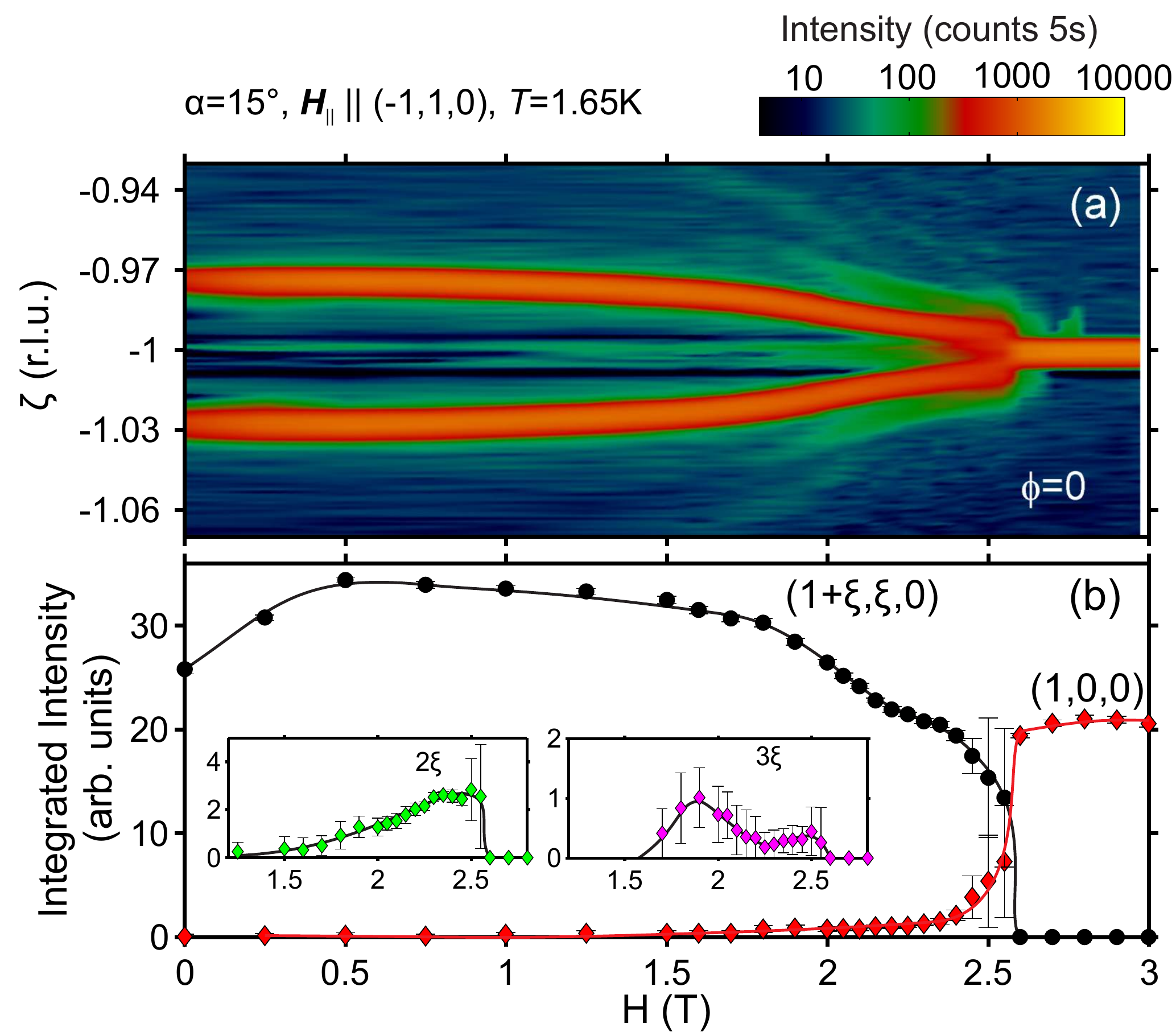}
\caption{Color online Neutron diffraction as function of magnetic field for $\alpha=15^{\circ}$ ((1,1,0) plane), $T=1.65\,{\rm K}$. Panel (a): Field dependence of the incommensurate satellite peaks around (1,0,0) for $\phi=0^{\circ}$. Panel (b): Field dependence of the integrated intensity of the satellite reflections at $\xi$ and the commensurate magnetic Bragg peak at (1,0,0), as obtained from Gaussian fits. The insets shows the magnetic field dependence of the intensity of the higher order reflections at $2\xi$ and $3\xi$. The lines serve as guide to the eye.}
\label{Fig_8}
\vspace{-0.02\textwidth}
\end{center}
\end{figure}

\begin{figure}
\begin{center}
\includegraphics[width=0.48\textwidth]{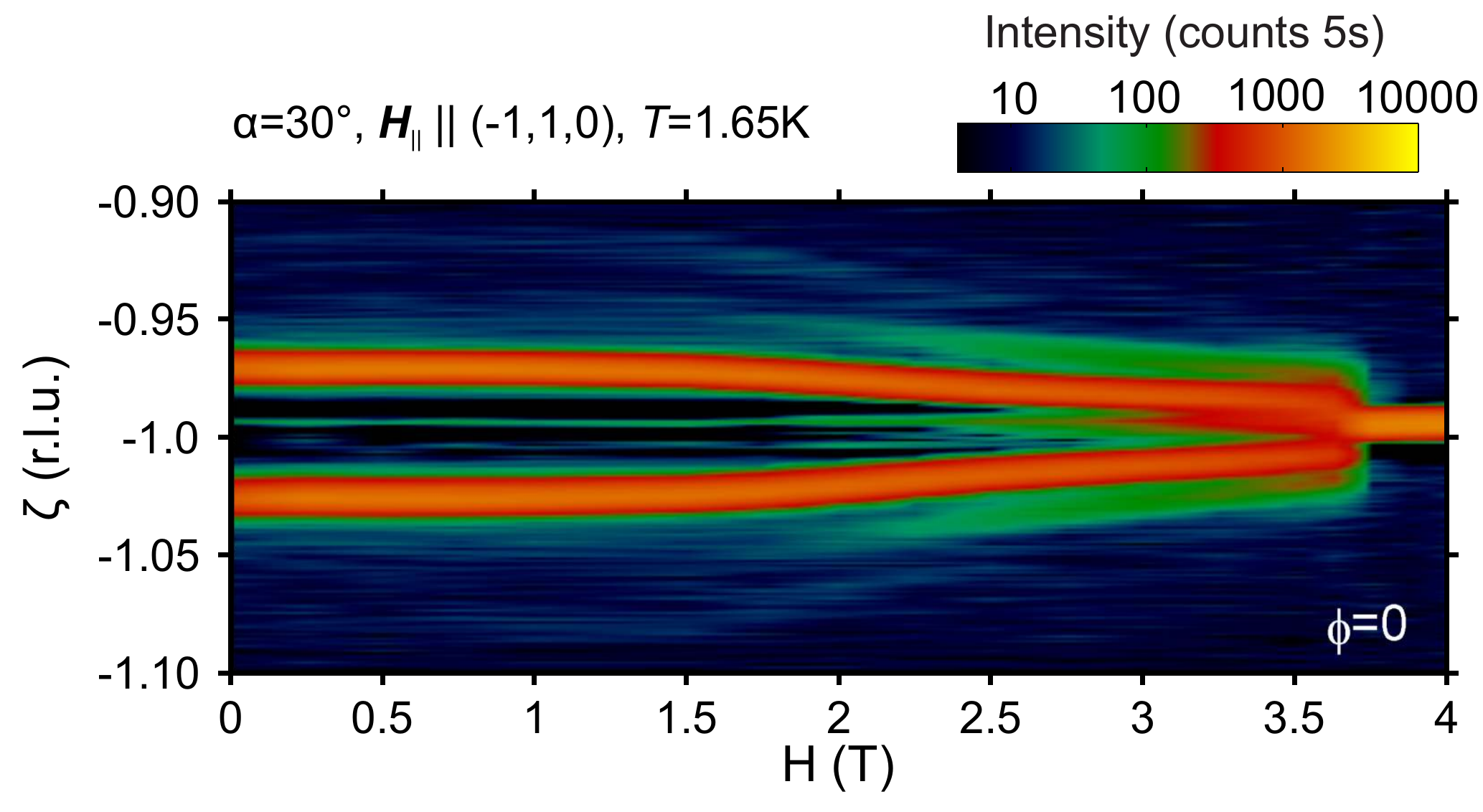}
\caption{Color online Magnetic field dependence of the satellite peaks around (1,0,0) for $\phi=0^{\circ}$, $\alpha=30^{\circ}$ ((1,1,0) plane) and $T=1.65\,{\rm K}$.}
\label{Fig_9}
\vspace{-0.02\textwidth}
\end{center}
\end{figure}

Neutron diffraction data for $\alpha=15^{\circ}$ confined in the (1,1,0) plane is shown in Fig.\,\,\ref{Fig_8}. Only the magnetic field dependence of only the domain at $\phi=0^{\circ}$ is shown. Panel (a) depicts the magnetic field dependence of the incommensurate satellite reflections around (1,0,0) while panel (b) shows the integrated intensity of $\xi$, the higher-order diffraction peaks at $2\xi$ and $3\xi$ and the commensurate peak at (1,0,0), as obtained from Gaussian fits.

As the in-plane component $\bm H_\|$ is oriented along the diagonal of the $(a,b)$-plane, no gradual rotation of the propagation vector is observed \cite{Zheludev:97b}. Similar to the previous setup, no flip of the propagation vector is observed for large $\alpha$. Again, for magnetic fields above $\approx 1.5\,{\rm T}$, higher order reflections at even and odd multiplies of $\xi$ are clearly seen (inset of panel (b)).  The field dependence of the odd higher order reflections at 3$\xi$ is quite remarkably: In contrast to the intensity observed at 2$\xi$, which linearly increases with field, the intensity of 3$\xi$ first increases and then shows a shallow dip until a second maximum is seen close to the I/C transition at $H_c= 2.5\,{\rm T}$. Note, that this feature of 3$\xi$ is not seen in the previous setup ((1,0,0) plane) due to the limited range of scans.

We finally dicuss the neutron diffraction data obtained for a tilted field of $\alpha=30^{\circ}$ confined in the (1,1,0) plane. Shown in the color map of Fig.\,\,\ref{Fig_9} is the magnetic field dependence of the domain at $\phi=0^{\circ}$. Similar to the previous setups with $\alpha=15^{\circ}$, higher order reflections at even and odd reflections at 2$\xi$ and 3$\xi$ smoothly appear above $H\approx 2\,{\rm T}$. The I/C transition is observed at a higher field of $H_c= 3.7\,{\rm T}$.

\subsubsection{Field dependence of the incommensurability parameter}

\begin{figure}
\begin{center}
\includegraphics[width=0.46\textwidth]{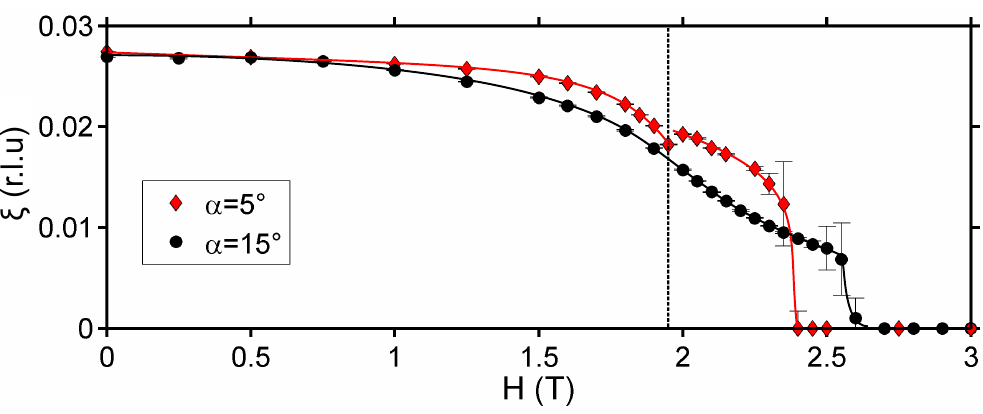}
\caption{Color online Magnetic field dependence of the incommensurability parameter $\xi$ for $\alpha=5^{\circ}$ and $\alpha=15^{\circ}$. The magnetic field was confined in the (1,1,0) plane. The solid lines serve as guide to the eye.}
\label{Fig_10}
\vspace{-0.02\textwidth}
\end{center}
\end{figure}

As remaining issue, we discuss the magnetic field dependence of the incommensurability parameter $\xi$ for different orientations of $\bm H$. Fig.\,\,\ref{Fig_10} shows the incommensurability parameter as obtained from the Gaussian fits for $\alpha=5^{\circ}$ and $\alpha=15^{\circ}$ with the magnetic field confined in the (1,1,0) plane. For these setups, no continuous rotation of the propagation vector ${\bm q}$ is hindering the analysis of the diffraction data. The position of opposite reflections at $\pm\xi$ has been averaged. Two salient features characterize the behaviour of $\xi(H)$: Firstly, for $\alpha=5^{\circ}$, a negative curvature of $\xi(H)$ is observed throughout the entire range of magnetic field. Secondly, at $H_1= 1.95\,{\rm T}$, a discontinuous jump of $\xi(H)$ to a slightly larger value is observed, corresponding to the phase transition from the soliton lattice to the AF-cone phase. For $H>H_1$ the data is consistent with a continuously divergent periodicity of the AF-cone phase. In contrast, for $\alpha=15^{\circ}$, no discontinuity of $\xi(H)$ is observed. At the same time, for magnetic fields corresponding to the emergence of odd and even higher harmonics, the curvature of $\xi(H)$ changes sign from negative to positive. Due to the smallness of $\xi$ close to the critical field, no reliable conclusion can be drawn whether $\xi(H)$ continuously diverges or shows a finite jump at $H_c$.

\section{IV. Discussion}

For an interpretation of our data, we first evaluate the phase diagram of Ba$_2$CuGe$_2$O$_7$ as function of the orientation of the magnetic field $\bm H$. We then continue with the phase diagrams in the $(H,T)$-plane for different $\alpha$, before we finally propose a tentative model of the different magnetic structures and their transitions.

\subsection{Phase transitions in canted magnetic fields}

\begin{figure}
\begin{center}
\includegraphics[width=0.48\textwidth]{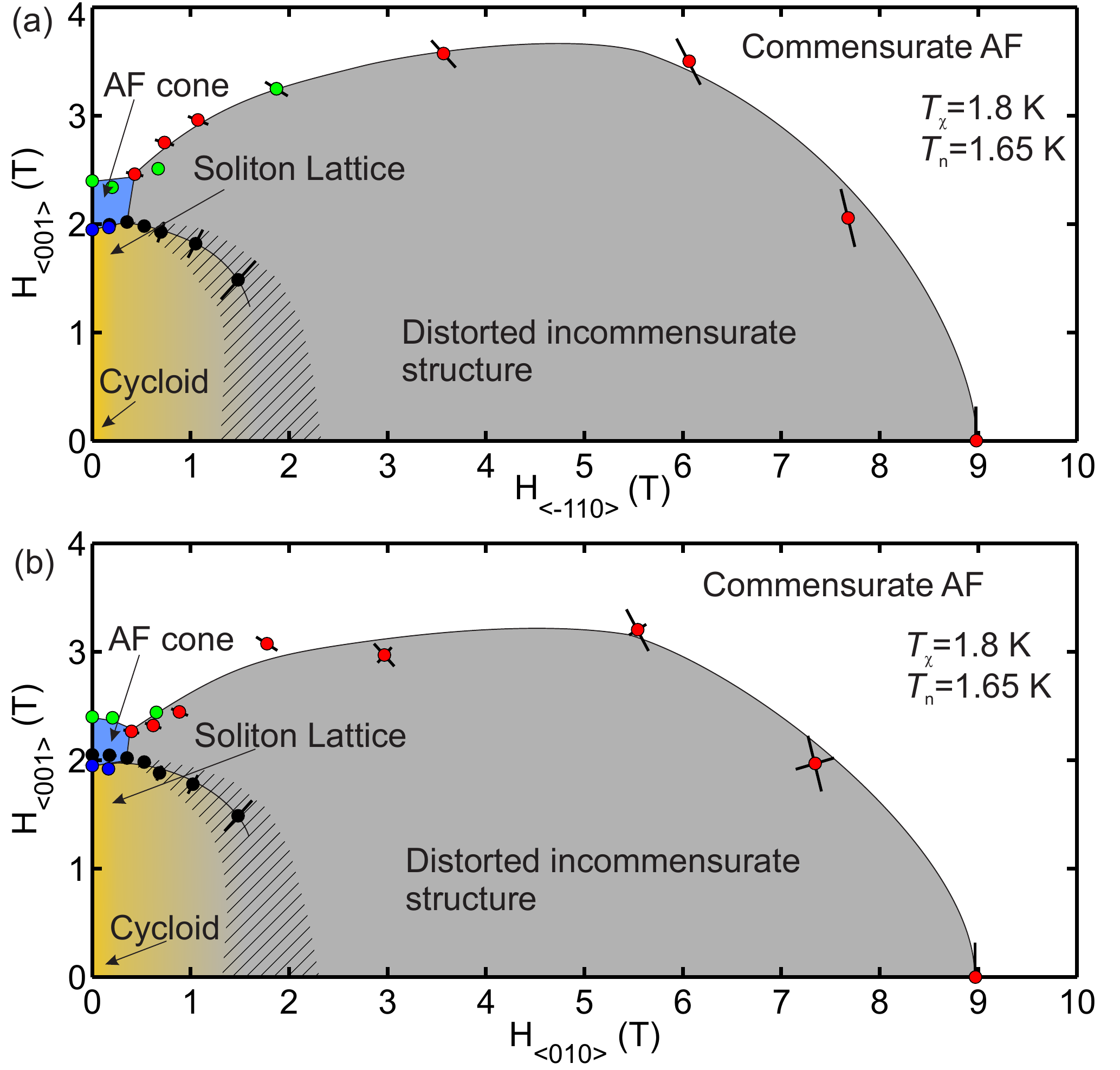}
\caption{Color online Proposed phase diagrams of Ba$_2$CuGe$_2$O$_7$ for magnetic field confined in the (1,1,0) plane (a) and the (1,0,0) plane (b). Green and blue markers indicate the I/C transition at $H_c$ and the phase transition from the soliton lattice to the AF-cone phase at $H_1$ as seen by neutron diffraction, respectively. Black points stand for $H_1$ as obtained by magnetic susceptibility. The hatched area thereby corresponds to the region where a broad crossover is observed. Red markers indicate the I/C transition at $H_c$, seen by magnetic susceptibility. Neutron data points have been taken at $T=1.65\,{\rm K}$, magnetic susceptibility at $T=1.8\,{\rm K}$, respectively. The lines serve as guide to the eye.}.
\label{Fig_11}
\vspace{-0.03\textwidth}
\end{center}
\end{figure}

Proposed magnetic phase diagrams of Ba$_2$CuGe$_2$O$_7$ are shown in Fig.\,\,\ref{Fig_11}. They summarize both neutron diffraction data and measurements of magnetic susceptibility, taken at $T_n=1.65\,{\rm K}$ and $T_{\chi}=1.8\,{\rm K}$, respectively. Panel (a) shows the phase diagram for magnetic field $\bm H$ confined in the (1,1,0) plane, panel (b) for the (1,0,0) plane. We summarize the salient features of the different phases and phase transitions:

(i) Apart from the continuous rotation of the propagation vector for low magnetic fields $H<0.5\,{\rm T}$ in agreement with Ref.\,\,\cite{Zheludev:97b}, virtually identical behaviour is seen for magnetic field $\bm H$ confined in both a (1,0,0) and a (1,1,0) plane.

(ii) In zero field, an {\it almost} AF cycloidal spin structure is observed: The ideal cycloidal spin structure is slightly distorded due to the KSEA interaction \cite{Zheludev:98bla, Zheludev:99}, giving rise to weak higher harmonics at $3\xi$. The cycloid significantly distorts to a lattice of isolated kinks (soliton lattice) for increasing magnetic field applied along the $c$-axis \cite{Zheludev:98PRB}, characterized by the increasing weight of higher harmonics at $3\xi$. 

(iii) The existence of the previously proposed AF-cone phase \cite{Muehlbauer:11} is confirmed consistently in both neutron scattering and measurements of magnetic susceptibility. The AF-cone phase is found to be stable for fields close to the I/C transition, closely aligned to the $c$-axis for $0^{\circ}\leq\alpha\lesssim 10^{\circ}$ for both the (1,0,0) and (1,1,0) plane. The small phase region of the AF-cone phase is shaded in blue in Fig.\,\,\ref{Fig_11}.

(iv) The transition from the soliton lattice to the AF-cone phase is characterized by a discontinuity of the propagation vector $\xi(H)$ at $H=H_1$. Higher order reflections at $3\xi$ are absent in the AF-cone phase. The AF-cone phase is further characterized by a relative flip of the propagation vector by $\pi/2$ \cite{Muehlbauer:11}. Neutron polarization confirmes that the commensurate component of the 2$k$ structure is aligned perpendicular to the in plane component of the magnetic field $\bm H_\|$. The plane of spin-rotation of the incommensurate component contains $\bm H_\|$ and is oriented perpendicular to the commensurate component. In contradiction to the interpretation of previous work \cite{Zheludev:98PRB}, no signature of the I/C transition is seen in magnetic susceptibility for the AF-cone phase.

(v) For $\alpha\gtrsim 10^{\circ}$ a crossover from the soliton lattice to a distorted incommensurate phase is seen in neutron diffraction and magnetization (area shaded in gray). The crossover to the distorted phase is characterized by the emergence of both odd and even multiplies of $\xi$ and accompagnied by a broad peak in $\rm d \bm M/\rm d \bm H$. The region of the crossover is indicated by the hatched area in Fig.\,\,\ref{Fig_11}. The width of the crossover thereby grows with increasing canting angle $\alpha$. In contrast to the AF-cone phase, a pronouned peak in $\rm d \bm M/\rm d \bm H$ characterizes the I/C transition for the distorted phase. The field dependence of the incommensurability parameter $\xi(H)$ furthermore suggests a discontinuous transition at $H=H_c$ rather than a continuously divergent periodicity.

(vi) The I/C transition can be continuously followed in magnetic suspectibility up to $H_c=9\,{\rm T}$ for magnetic field in the $(a,b)$-plane, both along the (1,0,0) and (1,1,0) direction ($\alpha=90^{\circ}$).

(vii) The finite magnetic susceptibility for $H>H_c$ indicates that the saturation field is not yet reached. The presence of the reflection at (1,0,0) is consistent with a commensurate AF phase (white area of  Fig.\,\,\ref{Fig_11}).

\subsection{$(H,T)$-Phase diagrams}

\begin{figure*}
\begin{center}
\includegraphics[width=0.98\textwidth]{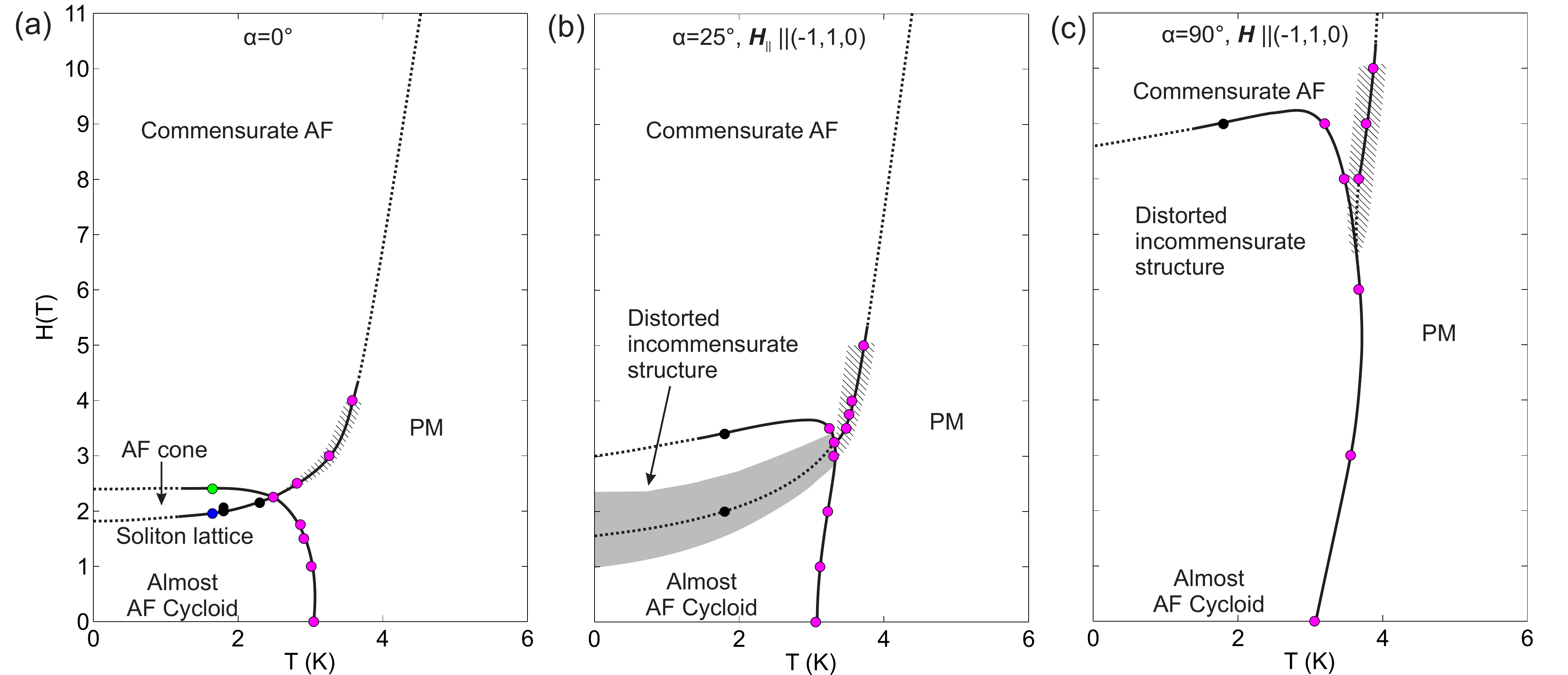}
\caption{Color online $(H,T)$-Phase diagram of Ba$_2$CuGe$_2$O$_7$. The same color code as for Fig.\,\,\ref{Fig_11} is used. Data points obtained by measurements of the specific heat are shown in magenta. Panel (a) shows the phase diagram for magnetic field aligned along the $c$-axis. Panel (b) shows the phase diagram for $\alpha=25^{\circ}$, confined in a (1,1,0) plane. Panel (c) shows the phase diagram for magnetic field along the (-1,1,0) direction ($\alpha=90^{\circ}$). Solid lines serve as guide to the eye. Dashed lines represent the assumed continuation of the phase transition lines, based on our model of the magnetic structure.}
\label{Fig_12}
\vspace{-0.02\textwidth}
\end{center}
\end{figure*}

The magnetic specific heat $C_{pm}$ of Ba$_2$CuGe$_2$O$_7$ is characterized by two separate features: A broad peak below $T=10\,{\rm K}$, centered at $T\sim 4.5\,{\rm K}$ and a sharp lambda anomaly at $T_N=3.2\,{\rm K}$. This behaviour is attributed to the quasi 2D character of Ba$_2$CuGe$_2$O$_7$. For a strictly two-dimensional system the Mermin-Wagner theorem dictates the absence of long-range order for any temperature $T>0$ and an exponentially diverging correlation length. A broad peak of the specific heat is found at $T/J\sim 0.7$. This feature is dominant for Ba$_2$CuGe$_2$O$_7$ as well, caused by short range antiferromagnetic 2D correlations in the basal $(a,b)$-plane, however with slightly reduced temperature. Any finite inter-plane coupling then leads to a phase transition to an ordered state for $T_c>0$, on top of the short range correlations, as observed for Ba$_2$CuGe$_2$O$_7$ at $T_N=3.2\,{\rm K}$. Stochastic series expansion quantum Monte Carlo simulations for a three-dimensional AF Heisenberg model on a cubic lattice with varying inter-plane coupling \cite{Sengupta:03} qualitatively reproduce the two peak structure in specific heat for the exchange ratio of Ba$_2$CuGe$_2$O$_7$ $\frac{J_{\perp}}{J_{\|}} \approx 2^{-5}$.

The transition at $T_N=3.2\,{\rm K}$ is characteristic of the spiral 3D long range order. The phase diagrams of Ba$_2$CuGe$_2$O$_7$ in the $(H,T)$-plane are summarized in Fig.\,\,\ref{Fig_12}, containing the data obtained in the neutron diffraction experiments, measurements of magnetization and specific heat. The same color code as for Fig.\,\,\ref{Fig_11} is used. For broadened peaks above $H_c$ (indicated by the hatched area), the maximum of the peak is defined as transition temperature. The dashed lines represent the assumed continuation of the phase transition lines, based on our model of the magnetic structure. Panel (a) shows the phase digram for the magnetic field $\bm H$ aligned along the $c$-axis ($\alpha=0^{\circ}$). The phase diagram is totally consistent with previously published data (Fig.\,5 of Ref.\,\,\cite{Zheludev:98PRB}). The previously called intermediate phase thereby precisely corresponds to the AF-cone phase. In addition, the model is consistent with a critical point at $T_c=2.5\,{\rm K}$ and $H_c=2.0\,{\rm T}$, explaining the absence of a lambda transition in specific heat at the critical magnetic field.

The $(H,T)$-phase diagram for tilted field is shown in panel (b) of Fig.\,\,\ref{Fig_12}. The magnetic field $\bm H$ is tilted in the (1,1,0) plane by $\alpha=25^{\circ}$. The gray area around the dashed horizontal transition line at $\approx 2\,{\rm T}$ denotes the crossover from the almost AF cycloidal phase at low-field to the distorted phase at high fields, as measured by magnetic susceptibility. Compared to the phase diagram obtained for magnetic field along the $c$-axis, the main differences are (i) a slightly increased critical field and (ii) the absence of the AF-cone phase. A similar trend is observed for the $H,T$-phase diagram for magnetic field parallel to the (1,-1,0) direction ($\alpha=90^{\circ}$). Here, the critical field is shifted to $\approx 9\,{\rm T}$, consistent with magnetic susceptibility. The crossover from the almost AF cycloidal phase to the distorted phase is completely smeared out and hence featureless in magnetic susceptibility.

For $H>H_c$ the commensurate AF to paramagnet transition generally appears as a broad feature in $C_{pm}$ which suggests a crossover instead of a transition. It is unclear whether the reason for this broadening is connected to the DM physics of Ba$_2$CuGe$_2$O$_7$ or caused by its weakly coupled 2D magnetic structure. The shift of specific heat from the background maximum, caused by the 2D-correlations, to the peak associated with the commensurate AF to paramagnet transition seen for high magnetic fields however indicates that the the DM physics migth be of minor importance in this regime of the phase diagram.

\subsection{Proposed magnetic structures}

\begin{figure*}
\begin{center}
\includegraphics[width=0.96\textwidth]{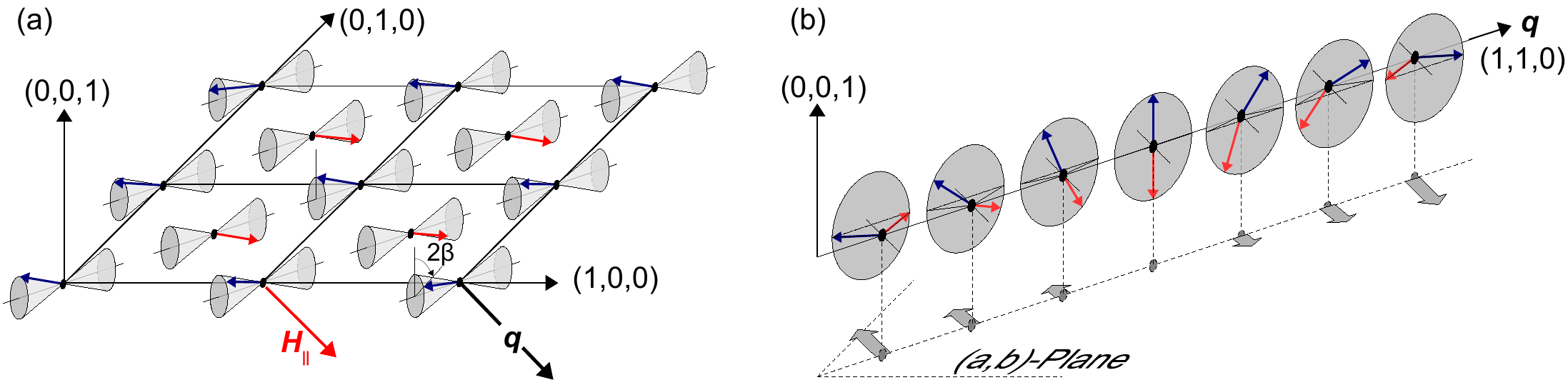}
\caption{Color online Magnetic structures of Ba$_2$CuGe$_2$O$_7$. Panel (a): Proposed structure of the incommensurate AF-cone phase. Panel (b): Zero field distortion of the almost AF cycloidal structure of Ba$_2$CuGe$_2$O$_7$ leading to a finite uniform magnetization for alternating periods of the cycloid. Shown is a projection of AF spins along the (-1,1,0) direction.}
\label{Fig_13}
\vspace{-0.02\textwidth}
\end{center}
\end{figure*}

Based on our experimental data, we finally propose a tentative model for the magnetic structures of Ba$_2$CuGe$_2$O$_7$. A schematic depiction of the almost AF cycloidal spin-structure has already been given in Fig.\,\,\ref{Fig_1}, panel (b). A sketch of the proposed AF-cone phase is given in Fig.\,\,\ref{Fig_13}, panel (a).

It has been found in previous neutron diffraction experiments \cite{Zheludev:97b} that sizable magnetic fields in the $(a,b)$-plane tend to rotate the plane of spin rotation of the almost AF cycloidal spin structure perpendicular to $\bm H_\|$, quantitatively explained by the interplay of Zeemann energy and the square crystal anisotropy. When a magnetic field is applied in the plane of spin rotation along the $c$-axis, the distortion and formation of a soliton lattice is hence the only way to respond for the almost cycloidal structure whilst conserving the plane of spin rotation. Note, that this distortion is symmetric with respect to the nodes of the soliton lattice, giving rise to higher order diffraction peaks at odd multiplies of the propagation vector. Experimental results for the intensity of the higher order diffraction peaks $\frac{I_{3\xi}}{I_{1\xi}}$ and the field dependence $\xi(H)$ agree well with an exact analytical solution \cite{Zheludev:97, Zheludev:98PRB}. In contrast, as already pointed out in Ref.\,\,\cite{Muehlbauer:11}, the AF-cone phase can take advantage of the Zeeman energy by canting its dominant AF-commensurate component, which is perpendicular to the $c$-axis. The deviation of the AF-cone phase from an ideal sinusoidal rotation is therefore less as compared to soliton lattice, explaining the systematic weakness of higher order reflections. This also provides a tentative explanation for the discontinuity seen at $H=H_1$, where $\xi(H)$ increases by $\approx 6\%$.

Our diffraction data is fully consistent with the model of the AF-cone phase as sketched in Fig.\,\,\ref{Fig_13}, panel (a: Neutron polarization confirms the perpendicular alignment of the commensurate component and the incommensurate rotating component. Our data further confirms a perpendicular orientation of the in-plane component of the magnetic field $\bm H_\|$ and the dominant commensurate component of the AF-cone phase, responsible for the flip of the propagation vector by $\pi/2$. Note, that the empiric rules for the rotation of plane-of-spin-rotation and propagation vector \cite{Zheludev:97b} seem to be valid for the AF-cone phase as well. We therefore assume that the AF-cone phase can freely rotate in the $(a,b)$-plane, due to a similar mechanism as for soliton lattice at lower fields. This behaviour is underscored by two experimental facts: Firstly, for $\bm H_\|$ along (0,1,0) the flip of the propagation vector by $\pi/2$ is seen {\it on top} of the continuous rotation. Secondly, a virtually identical behaviour has been found for magnetic field confined in the (1,0,0) and (1,1,0) plane.

The AF-cone phase was largely missed in most theoretical descriptions of Ba$_2$CuGe$_2$O$_7$ \cite{Zheludev:97PB,Zheludev:97b, Zheludev:98PRL,Zheludev:98PRB, Bogdanov:99, Bogdanov:02, Bogdanov:04}. In contrast, using a continuum field theory in the form of a non-linear $\sigma$ model, Chovan {\it et al.} \cite{Chovan:02} have indeed suggested a magnetic structure in the form of a conical spiral that nutates around the (110) direction, containing all three components of the magnetization. A numerical relaxation method was used on a periodic 1D grid. This conical spiral was shown to be stable in a field range between $1.7\,\rm{T}$ and $2.9\,\rm{T}$ for magnetic field aligned along the $c$-direction and degenerates to a spin-flop phase at the upper critical field, naturally exlaining the absence of a signature in susceptibility. Furthermore, the model by Chovan {\it et al.} qualitatively reproduces the field dependence of the incommensurability parameter. However, further theoretical input is needed to understand the stabilization of the AF-cone, in particular its sensitivity to in plane components of the magnetic field and the nature of the phase transitions. Due to the softness of the original model close to the I/C transition, additional terms in the Hamiltonian need not to be strong. Two possible candidates have already pointed out by \cite{Muehlbauer:11}. Firstly, lattice effects and secondly the sign alternating component of the DM vector ${\bm D}_z$ which has often been ignored, giving rise to $D_z(S^{(x)}_1S^{(y)}_2-S^{(y)}_1S^{(x)}_2)$. The presence of ${\bm D}_z$ favours a commensurate weak-ferromagnetic canting of all spins in one direction \cite{Bogdanov:04} as for example in the similar K$_2$V$_3$O$_8$ system \cite{Lumsden2001}.

In the presence of an almost cycloidal structure with its plane of spin rotation containing the $c$-axis as in Ba$_2$CuGe$_2$O$_7$, ${\bm D}_z$ will mainly affect those spins of the cycloid lying closest to the $(a,b)$-plane. The result is a ferromagnetic canting of spins which is sign alternating for alternating periods of the cycloid. The ferromagnetic moment is aligned in the $(a,b)$-plane. Such structures have been predicted by theory \cite{Bogdanov:02}. A schematic depiction is given in Fig.\,\,\ref{Fig_13}, panel (b), showing a projection of AF spins along the (-1,1,0) direction. Note, that in the presence of a magnetic field component in the $(a,b)$-plane, this would lead to asymmetric distortion of the cyloidal structure, compressing next-nearest periods of the cycloid. Such asymmetric distortion gives rise to {\it even}-order harmonics. Indeed, such even harmonics are observed for increasing magnetic field components in the $(a,b)$-plane of Ba$_2$CuGe$_2$O$_7$. We therefore speculate that ${\bm D}_z$ in fact plays an important role for the crossover from the soliton lattice to the non-sinusoidal distorted structure for tilted magnetic field. Our assumptions are underscored by the observation of an unexpected I/C transition for magnetic field in the basal $(a,b)$-plane. This I/C transition is clearly different from the I/C transition, observed for magnetic field aligned parallel to the $c$-axis. For the latter, a sharp signature in magnetic susceptibility is systematically absent. Moreover, the curvature of $\xi(H_c)$ close to the transition has opposite sign.

\section{V. Conclusion}
In our work, a detailed experimental picture of the different phases and phase transitions of  Ba$_2$CuGe$_2$O$_7$ has been obtained as function of the orientation of the magnetic field $\bm H$. Using neutron diffraction in combination with measurements of magnetic susceptibility and specific heat we confirm the existence of the previously proposed incommensurate AF-cone phase \cite{Muehlbauer:11}. We show that the AF-cone phase is only stable for a small magnetic field region close to the $c$-axis for magnetic field in both a (1,1,0) and (1,0,0) plane. For large angles enclosed by $\bm H$ and the $c$-axis, a complexely distorted non-sinusoidal magnetic structure has recently been observed \cite{Muehlbauer:11}. We systematically examine the behaviour of the distorted phase and find virtually identical behaviour for magnetic field applied in the (1,1,0) and (1,0,0) plane. We show that the critical field $H_c$ systematically increases for larger canting. In fact, measurements of magnetic susceptibility and specific heat indicate the existence of a unexpected I/C transition for magnetic fields applied in the basal $(a,b)$-plane at $H_c\approx 9\,\,{\rm T}$. The appearance of odd and even harmonics of $\bm q$ at $2\xi$ and $3\xi$ are consistent with a non-planar, distorted non-sinusoidal magnetic structure, however the detailed structure of the distorted phase still remains unresolved. We have furthermore identified the staggered DM vector ${\bm D_z}$ as a key element for understanding the variety of magnetic structures in Ba$_2$CuGe$_2$O$_7$. For future experimental investigations, we plan to use polarized neutron diffraction as well as small-angle-neutron scattering. A further understanding of the mechanism responsible for the stabilization of the different spin-structures of Ba$_2$CuGe$_2$O$_7$ will clearly require new theoretical input.

\section{VI. Acknowledgments}

We wish to thank K. Conder, J. Kohlbrecher, C. Niedermayer, J. Gavilano, M. Laver, C. Pfleiderer, P. B\"oni, W. Lorenz, U. R\"ossler, M. Garst and A. Rosch for support and stimulating discussions. Technical support and help from M. Zolliker, P. Fouilloux, M. Bartkowiak and R. Frison is gratefully acknowledged. This work is partially supported through Project 6 of MANEP, Swiss National Science Foundation.


\begin{thebibliography}{32}
\expandafter\ifx\csname natexlab\endcsname\relax\def\natexlab#1{#1}\fi
\expandafter\ifx\csname bibnamefont\endcsname\relax
  \def\bibnamefont#1{#1}\fi
\expandafter\ifx\csname bibfnamefont\endcsname\relax
  \def\bibfnamefont#1{#1}\fi
\expandafter\ifx\csname citenamefont\endcsname\relax
  \def\citenamefont#1{#1}\fi
\expandafter\ifx\csname url\endcsname\relax
  \def\url#1{\texttt{#1}}\fi
\expandafter\ifx\csname urlprefix\endcsname\relax\def\urlprefix{URL }\fi
\providecommand{\bibinfo}[2]{#2}
\providecommand{\eprint}[2][]{\url{#2}}

\bibitem[{\citenamefont{M{\"u}hlbauer et~al.}(2009)\citenamefont{M{\"u}hlbauer,
  Binz, Jonietz, Pfleiderer, B{\"o}ni, Rosch, Neubauer, and
  Georgii}}]{Muehlbauer:09b}
\bibinfo{author}{\bibfnamefont{S.}~\bibnamefont{M{\"u}hlbauer}},
  \bibinfo{author}{\bibfnamefont{B.}~\bibnamefont{Binz}},
  \bibinfo{author}{\bibfnamefont{F.}~\bibnamefont{Jonietz}},
  \bibinfo{author}{\bibfnamefont{C.}~\bibnamefont{Pfleiderer}},
  \bibinfo{author}{\bibfnamefont{P.}~\bibnamefont{B{\"o}ni}},
  \bibinfo{author}{\bibfnamefont{A.}~\bibnamefont{Rosch}},
  \bibinfo{author}{\bibfnamefont{A.}~\bibnamefont{Neubauer}}, \bibnamefont{and}
  \bibinfo{author}{\bibfnamefont{R.}~\bibnamefont{Georgii}},
  \bibinfo{journal}{Science} \textbf{\bibinfo{volume}{323}},
  \bibinfo{pages}{5916} (\bibinfo{year}{2009}).

\bibitem[{\citenamefont{M\"unzer et~al.}(2010)\citenamefont{M\"unzer, Neubauer,
  Adams, M\"uhlbauer, Franz, Jonietz, Georgii, B\"oni, Pedersen, Schmidt
  et~al.}}]{Muenzer:09}
\bibinfo{author}{\bibfnamefont{W.}~\bibnamefont{M\"unzer}},
  \bibinfo{author}{\bibfnamefont{A.}~\bibnamefont{Neubauer}},
  \bibinfo{author}{\bibfnamefont{T.}~\bibnamefont{Adams}},
  \bibinfo{author}{\bibfnamefont{S.}~\bibnamefont{M\"uhlbauer}},
  \bibinfo{author}{\bibfnamefont{C.}~\bibnamefont{Franz}},
  \bibinfo{author}{\bibfnamefont{F.}~\bibnamefont{Jonietz}},
  \bibinfo{author}{\bibfnamefont{R.}~\bibnamefont{Georgii}},
  \bibinfo{author}{\bibfnamefont{P.}~\bibnamefont{B\"oni}},
  \bibinfo{author}{\bibfnamefont{B.}~\bibnamefont{Pedersen}},
  \bibinfo{author}{\bibfnamefont{M.}~\bibnamefont{Schmidt}},
  \bibnamefont{et~al.}, \bibinfo{journal}{Phys. Rev. B}
  \textbf{\bibinfo{volume}{81}}, \bibinfo{pages}{041203}
  (\bibinfo{year}{2010}).

\bibitem[{\citenamefont{Yu et~al.}(2011)\citenamefont{Yu, Kanazawa, Onose,
  Kimoto, Zhang, Isiwata, Matsui, and Tokura}}]{Yu:11}
\bibinfo{author}{\bibfnamefont{X.~Z.} \bibnamefont{Yu}},
  \bibinfo{author}{\bibfnamefont{N.}~\bibnamefont{Kanazawa}},
  \bibinfo{author}{\bibfnamefont{Y.}~\bibnamefont{Onose}},
  \bibinfo{author}{\bibfnamefont{K.}~\bibnamefont{Kimoto}},
  \bibinfo{author}{\bibfnamefont{W.~Z.} \bibnamefont{Zhang}},
  \bibinfo{author}{\bibfnamefont{S.}~\bibnamefont{Isiwata}},
  \bibinfo{author}{\bibfnamefont{Y.}~\bibnamefont{Matsui}}, \bibnamefont{and}
  \bibinfo{author}{\bibfnamefont{Y.}~\bibnamefont{Tokura}},
  \bibinfo{journal}{Nature Materials} \textbf{\bibinfo{volume}{10}},
  \bibinfo{pages}{106} (\bibinfo{year}{2011}).

\bibitem[{\citenamefont{Yu et~al.}(1999)\citenamefont{Yu, Onose, Kanazawa,
  Park, Han, Matsui, Nagaosa, and Tokura}}]{Yu:10}
\bibinfo{author}{\bibfnamefont{X.~Z.} \bibnamefont{Yu}},
  \bibinfo{author}{\bibfnamefont{Y.}~\bibnamefont{Onose}},
  \bibinfo{author}{\bibfnamefont{N.}~\bibnamefont{Kanazawa}},
  \bibinfo{author}{\bibfnamefont{J.~H.} \bibnamefont{Park}},
  \bibinfo{author}{\bibfnamefont{J.~H.} \bibnamefont{Han}},
  \bibinfo{author}{\bibfnamefont{Y.}~\bibnamefont{Matsui}},
  \bibinfo{author}{\bibfnamefont{N.}~\bibnamefont{Nagaosa}}, \bibnamefont{and}
  \bibinfo{author}{\bibfnamefont{Y.}~\bibnamefont{Tokura}},
  \bibinfo{journal}{Nature} \textbf{\bibinfo{volume}{25}}, \bibinfo{pages}{76}
  (\bibinfo{year}{1999}).

\bibitem[{\citenamefont{Adams et~al.}(2011)\citenamefont{Adams, M\"uhlbauer,
  Pfleiderer, Jonietz, Bauer, Neubauer, Georgii, B\"oni, Keiderling, Everschor
  et~al.}}]{Adams:11}
\bibinfo{author}{\bibfnamefont{T.}~\bibnamefont{Adams}},
  \bibinfo{author}{\bibfnamefont{S.}~\bibnamefont{M\"uhlbauer}},
  \bibinfo{author}{\bibfnamefont{C.}~\bibnamefont{Pfleiderer}},
  \bibinfo{author}{\bibfnamefont{F.}~\bibnamefont{Jonietz}},
  \bibinfo{author}{\bibfnamefont{A.}~\bibnamefont{Bauer}},
  \bibinfo{author}{\bibfnamefont{A.}~\bibnamefont{Neubauer}},
  \bibinfo{author}{\bibfnamefont{R.}~\bibnamefont{Georgii}},
  \bibinfo{author}{\bibfnamefont{P.}~\bibnamefont{B\"oni}},
  \bibinfo{author}{\bibfnamefont{U.}~\bibnamefont{Keiderling}},
  \bibinfo{author}{\bibfnamefont{K.}~\bibnamefont{Everschor}},
  \bibnamefont{et~al.}, \bibinfo{journal}{arXiv:1107.0993}
  (\bibinfo{year}{2011}).

\bibitem[{\citenamefont{Mostovoy}(2006)}]{Mostovoy:06}
\bibinfo{author}{\bibfnamefont{M.}~\bibnamefont{Mostovoy}},
  \bibinfo{journal}{Phys. Rev. Lett.} \textbf{\bibinfo{volume}{96}},
  \bibinfo{pages}{067601} (\bibinfo{year}{2006}).

\bibitem[{\citenamefont{Sergienko and Dagotto}(2006)}]{Sergienko:06}
\bibinfo{author}{\bibfnamefont{I.~A.} \bibnamefont{Sergienko}}
  \bibnamefont{and} \bibinfo{author}{\bibfnamefont{E.}~\bibnamefont{Dagotto}},
  \bibinfo{journal}{Phys. Rev. B} \textbf{\bibinfo{volume}{73}},
  \bibinfo{pages}{094434} (\bibinfo{year}{2006}).

\bibitem[{\citenamefont{Kimura et~al.}(2005)\citenamefont{Kimura, Lawes, and
  Ramirez}}]{kimura:05}
\bibinfo{author}{\bibfnamefont{T.}~\bibnamefont{Kimura}},
  \bibinfo{author}{\bibfnamefont{G.}~\bibnamefont{Lawes}}, \bibnamefont{and}
  \bibinfo{author}{\bibfnamefont{A.~P.} \bibnamefont{Ramirez}},
  \bibinfo{journal}{Phys. Rev. Lett.} \textbf{\bibinfo{volume}{94}},
  \bibinfo{pages}{137201} (\bibinfo{year}{2005}).

\bibitem[{\citenamefont{Kenzelmann et~al.}(2005)}]{Kenzelmann:05}
\bibinfo{author}{\bibfnamefont{M.}~\bibnamefont{Kenzelmann}}
  \bibnamefont{et~al.}, \bibinfo{journal}{Phys. Rev. Lett.}
  \textbf{\bibinfo{volume}{95}}, \bibinfo{eid}{087206} (\bibinfo{year}{2005}).

\bibitem[{\citenamefont{Lawes et~al.}(2005)}]{Lawes:05}
\bibinfo{author}{\bibfnamefont{G.}~\bibnamefont{Lawes}} \bibnamefont{et~al.},
  \bibinfo{journal}{Phys. Rev. Lett.} \textbf{\bibinfo{volume}{95}},
  \bibinfo{pages}{087205} (\bibinfo{year}{2005}).

\bibitem[{\citenamefont{Dzyaloshinskii}(1958)}]{Dzyaloshinskii:58}
\bibinfo{author}{\bibfnamefont{I.~E.} \bibnamefont{Dzyaloshinskii}},
  \bibinfo{journal}{J. Phys. Chem Solids} \textbf{\bibinfo{volume}{4}},
  \bibinfo{pages}{241} (\bibinfo{year}{1958}).

\bibitem[{\citenamefont{Moriya}(1960)}]{Moriya:60}
\bibinfo{author}{\bibfnamefont{T.}~\bibnamefont{Moriya}},
  \bibinfo{journal}{Phys. Rev.} \textbf{\bibinfo{volume}{120}},
  \bibinfo{pages}{91} (\bibinfo{year}{1960}).

\bibitem[{\citenamefont{Zheludev et~al.}(1996)\citenamefont{Zheludev, Shirane,
  Sasago, Kiode, and Uchinokura}}]{Zheludev:96}
\bibinfo{author}{\bibfnamefont{A.}~\bibnamefont{Zheludev}},
  \bibinfo{author}{\bibfnamefont{G.}~\bibnamefont{Shirane}},
  \bibinfo{author}{\bibfnamefont{Y.}~\bibnamefont{Sasago}},
  \bibinfo{author}{\bibfnamefont{N.}~\bibnamefont{Kiode}}, \bibnamefont{and}
  \bibinfo{author}{\bibfnamefont{K.}~\bibnamefont{Uchinokura}},
  \bibinfo{journal}{Phys. Rev. B} \textbf{\bibinfo{volume}{54}},
  \bibinfo{pages}{15163} (\bibinfo{year}{1996}).

\bibitem[{\citenamefont{Yi et~al.}(2008)\citenamefont{Yi, Choi, Lee, and
  S.-W-Cheong}}]{Yi:08}
\bibinfo{author}{\bibfnamefont{H.~T.} \bibnamefont{Yi}},
  \bibinfo{author}{\bibfnamefont{Y.~J.} \bibnamefont{Choi}},
  \bibinfo{author}{\bibfnamefont{S.}~\bibnamefont{Lee}}, \bibnamefont{and}
  \bibinfo{author}{\bibnamefont{S.-W-Cheong}}, \bibinfo{journal}{Applied
  Physics Letters} \textbf{\bibinfo{volume}{92}}, \bibinfo{pages}{212904}
  (\bibinfo{year}{2008}).

\bibitem[{\citenamefont{Zheludev et~al.}(2003)\citenamefont{Zheludev, Sato,
  Masuda, Uchinokura, Shirane, and Roessli}}]{Zheludev:03}
\bibinfo{author}{\bibfnamefont{A.}~\bibnamefont{Zheludev}},
  \bibinfo{author}{\bibfnamefont{T.}~\bibnamefont{Sato}},
  \bibinfo{author}{\bibfnamefont{T.}~\bibnamefont{Masuda}},
  \bibinfo{author}{\bibfnamefont{K.}~\bibnamefont{Uchinokura}},
  \bibinfo{author}{\bibfnamefont{G.}~\bibnamefont{Shirane}}, \bibnamefont{and}
  \bibinfo{author}{\bibfnamefont{B.}~\bibnamefont{Roessli}},
  \bibinfo{journal}{Phys. Rev. B} \textbf{\bibinfo{volume}{68}},
  \bibinfo{pages}{024428} (\bibinfo{year}{2003}).

\bibitem[{\citenamefont{Sato et~al.}(2003)\citenamefont{Sato, Matsuda, and
  Uchinokura}}]{Sato:03}
\bibinfo{author}{\bibfnamefont{T.}~\bibnamefont{Sato}},
  \bibinfo{author}{\bibfnamefont{T.}~\bibnamefont{Matsuda}}, \bibnamefont{and}
  \bibinfo{author}{\bibfnamefont{K.}~\bibnamefont{Uchinokura}},
  \bibinfo{journal}{Physica B} \textbf{\bibinfo{volume}{329}},
  \bibinfo{pages}{880} (\bibinfo{year}{2003}).

\bibitem[{\citenamefont{Bogdanov et~al.}(2002)\citenamefont{Bogdanov,
  R\"o\ss{}ler, Wolf, and M\"uller}}]{Bogdanov:02}
\bibinfo{author}{\bibfnamefont{A.~N.} \bibnamefont{Bogdanov}},
  \bibinfo{author}{\bibfnamefont{U.~K.} \bibnamefont{R\"o\ss{}ler}},
  \bibinfo{author}{\bibfnamefont{M.}~\bibnamefont{Wolf}}, \bibnamefont{and}
  \bibinfo{author}{\bibfnamefont{K.-H.} \bibnamefont{M\"uller}},
  \bibinfo{journal}{Phys. Rev. B} \textbf{\bibinfo{volume}{66}},
  \bibinfo{pages}{214410} (\bibinfo{year}{2002}).

\bibitem[{\citenamefont{M\"uhlbauer et~al.}(2011)\citenamefont{M\"uhlbauer,
  Gvasaliya, Pomjakushina, and Zheludev}}]{Muehlbauer:11}
\bibinfo{author}{\bibfnamefont{S.}~\bibnamefont{M\"uhlbauer}},
  \bibinfo{author}{\bibfnamefont{S.~N.} \bibnamefont{Gvasaliya}},
  \bibinfo{author}{\bibfnamefont{E.}~\bibnamefont{Pomjakushina}},
  \bibnamefont{and} \bibinfo{author}{\bibfnamefont{A.}~\bibnamefont{Zheludev}},
  \bibinfo{journal}{Phys. Rev. B (R)} \textbf{\bibinfo{volume}{84}},
  \bibinfo{pages}{180406} (\bibinfo{year}{2011}),
  \urlprefix\url{http://link.aps.org/doi/10.1103/PhysRevB.84.180406}.

\bibitem[{\citenamefont{Zheludev
  et~al.}(1997{\natexlab{a}})\citenamefont{Zheludev, Maslov, Shirane, Sasago,
  Koide, Uchinokura, Tennant, and Nagler}}]{Zheludev:97b}
\bibinfo{author}{\bibfnamefont{A.}~\bibnamefont{Zheludev}},
  \bibinfo{author}{\bibfnamefont{S.}~\bibnamefont{Maslov}},
  \bibinfo{author}{\bibfnamefont{G.}~\bibnamefont{Shirane}},
  \bibinfo{author}{\bibfnamefont{Y.}~\bibnamefont{Sasago}},
  \bibinfo{author}{\bibfnamefont{N.}~\bibnamefont{Koide}},
  \bibinfo{author}{\bibfnamefont{K.}~\bibnamefont{Uchinokura}},
  \bibinfo{author}{\bibfnamefont{D.~A.} \bibnamefont{Tennant}},
  \bibnamefont{and} \bibinfo{author}{\bibfnamefont{S.~E.}
  \bibnamefont{Nagler}}, \bibinfo{journal}{Phys. Rev. B}
  \textbf{\bibinfo{volume}{56}}, \bibinfo{pages}{14006}
  (\bibinfo{year}{1997}{\natexlab{a}}).

\bibitem[{\citenamefont{Zheludev et~al.}(1999)\citenamefont{Zheludev, Maslov,
  Shirane, Tsukada, Masuda, Uchinokura, Zaliznyak, Erwin, and
  Regnault}}]{Zheludev:99}
\bibinfo{author}{\bibfnamefont{A.}~\bibnamefont{Zheludev}},
  \bibinfo{author}{\bibfnamefont{S.}~\bibnamefont{Maslov}},
  \bibinfo{author}{\bibfnamefont{G.}~\bibnamefont{Shirane}},
  \bibinfo{author}{\bibfnamefont{I.}~\bibnamefont{Tsukada}},
  \bibinfo{author}{\bibfnamefont{T.}~\bibnamefont{Masuda}},
  \bibinfo{author}{\bibfnamefont{K.}~\bibnamefont{Uchinokura}},
  \bibinfo{author}{\bibfnamefont{I.}~\bibnamefont{Zaliznyak}},
  \bibinfo{author}{\bibfnamefont{R.}~\bibnamefont{Erwin}}, \bibnamefont{and}
  \bibinfo{author}{\bibfnamefont{L.~P.} \bibnamefont{Regnault}},
  \bibinfo{journal}{Phys. Rev. B} \textbf{\bibinfo{volume}{59}},
  \bibinfo{pages}{11432} (\bibinfo{year}{1999}).

\bibitem[{\citenamefont{Zheludev
  et~al.}(1997{\natexlab{b}})\citenamefont{Zheludev, Maslov, Shirane, Sasago,
  Koide, and Uchinokura}}]{Zheludev:97}
\bibinfo{author}{\bibfnamefont{A.}~\bibnamefont{Zheludev}},
  \bibinfo{author}{\bibfnamefont{S.}~\bibnamefont{Maslov}},
  \bibinfo{author}{\bibfnamefont{G.}~\bibnamefont{Shirane}},
  \bibinfo{author}{\bibfnamefont{Y.}~\bibnamefont{Sasago}},
  \bibinfo{author}{\bibfnamefont{N.}~\bibnamefont{Koide}}, \bibnamefont{and}
  \bibinfo{author}{\bibfnamefont{K.}~\bibnamefont{Uchinokura}},
  \bibinfo{journal}{Phys. Rev. Lett.} \textbf{\bibinfo{volume}{78}},
  \bibinfo{pages}{4857} (\bibinfo{year}{1997}{\natexlab{b}}).

\bibitem[{\citenamefont{Zheludev
  et~al.}(1997{\natexlab{c}})\citenamefont{Zheludev, Shirane, Sasago, Koide,
  and Uchinokura}}]{Zheludev:97PB}
\bibinfo{author}{\bibfnamefont{A.}~\bibnamefont{Zheludev}},
  \bibinfo{author}{\bibfnamefont{G.}~\bibnamefont{Shirane}},
  \bibinfo{author}{\bibfnamefont{Y.}~\bibnamefont{Sasago}},
  \bibinfo{author}{\bibfnamefont{N.}~\bibnamefont{Koide}}, \bibnamefont{and}
  \bibinfo{author}{\bibfnamefont{K.}~\bibnamefont{Uchinokura}},
  \bibinfo{journal}{Physica B: Condensed Matter}
  \textbf{\bibinfo{volume}{234–236}}, \bibinfo{pages}{546 }
  (\bibinfo{year}{1997}{\natexlab{c}}), ISSN \bibinfo{issn}{0921-4526},
  \bibinfo{note}{proceedings of the First European Conference on Neutron
  Scattering}.

\bibitem[{\citenamefont{Zheludev
  et~al.}(1998{\natexlab{a}})\citenamefont{Zheludev, Maslov, Tsukada,
  Zaliznyak, Regnault, Masuda, Uchinokura, Erwin, and
  Shirane}}]{Zheludev:98bla}
\bibinfo{author}{\bibfnamefont{A.}~\bibnamefont{Zheludev}},
  \bibinfo{author}{\bibfnamefont{S.}~\bibnamefont{Maslov}},
  \bibinfo{author}{\bibfnamefont{I.}~\bibnamefont{Tsukada}},
  \bibinfo{author}{\bibfnamefont{I.}~\bibnamefont{Zaliznyak}},
  \bibinfo{author}{\bibfnamefont{L.~P.} \bibnamefont{Regnault}},
  \bibinfo{author}{\bibfnamefont{T.}~\bibnamefont{Masuda}},
  \bibinfo{author}{\bibfnamefont{K.}~\bibnamefont{Uchinokura}},
  \bibinfo{author}{\bibfnamefont{R.}~\bibnamefont{Erwin}}, \bibnamefont{and}
  \bibinfo{author}{\bibfnamefont{G.}~\bibnamefont{Shirane}},
  \bibinfo{journal}{Phys. Rev. Lett.} \textbf{\bibinfo{volume}{81}},
  \bibinfo{pages}{5410} (\bibinfo{year}{1998}{\natexlab{a}}),
  \urlprefix\url{http://link.aps.org/doi/10.1103/PhysRevLett.81.5410}.

\bibitem[{\citenamefont{Murakawa et~al.}(2009)\citenamefont{Murakawa, Onose,
  and Tokura}}]{Murakawa:09}
\bibinfo{author}{\bibfnamefont{H.}~\bibnamefont{Murakawa}},
  \bibinfo{author}{\bibfnamefont{Y.}~\bibnamefont{Onose}}, \bibnamefont{and}
  \bibinfo{author}{\bibfnamefont{Y.}~\bibnamefont{Tokura}},
  \bibinfo{journal}{Phys. Rev. Lett.} \textbf{\bibinfo{volume}{103}},
  \bibinfo{pages}{147201} (\bibinfo{year}{2009}),
  \urlprefix\url{http://link.aps.org/doi/10.1103/PhysRevLett.103.147201}.

\bibitem[{\citenamefont{Semadeni et~al.}(2001)\citenamefont{Semadeni, Roessli,
  and B{\"o}ni}}]{semadeni:01}
\bibinfo{author}{\bibfnamefont{F.}~\bibnamefont{Semadeni}},
  \bibinfo{author}{\bibfnamefont{B.}~\bibnamefont{Roessli}}, \bibnamefont{and}
  \bibinfo{author}{\bibfnamefont{P.}~\bibnamefont{B{\"o}ni}},
  \bibinfo{journal}{Physica B} \textbf{\bibinfo{volume}{297}},
  \bibinfo{pages}{152} (\bibinfo{year}{2001}).

\bibitem[{\citenamefont{Zheludev
  et~al.}(1998{\natexlab{b}})\citenamefont{Zheludev, Maslov, Shirane, Sasago,
  Koide, and Uchinokura}}]{Zheludev:98PRB}
\bibinfo{author}{\bibfnamefont{A.}~\bibnamefont{Zheludev}},
  \bibinfo{author}{\bibfnamefont{S.}~\bibnamefont{Maslov}},
  \bibinfo{author}{\bibfnamefont{G.}~\bibnamefont{Shirane}},
  \bibinfo{author}{\bibfnamefont{Y.}~\bibnamefont{Sasago}},
  \bibinfo{author}{\bibfnamefont{N.}~\bibnamefont{Koide}}, \bibnamefont{and}
  \bibinfo{author}{\bibfnamefont{K.}~\bibnamefont{Uchinokura}},
  \bibinfo{journal}{Phys. Rev. B} \textbf{\bibinfo{volume}{57}},
  \bibinfo{pages}{2968} (\bibinfo{year}{1998}{\natexlab{b}}).

\bibitem[{\citenamefont{Sengupta et~al.}(2003)\citenamefont{Sengupta, Sandvik,
  and Singh}}]{Sengupta:03}
\bibinfo{author}{\bibfnamefont{P.}~\bibnamefont{Sengupta}},
  \bibinfo{author}{\bibfnamefont{A.~W.} \bibnamefont{Sandvik}},
  \bibnamefont{and} \bibinfo{author}{\bibfnamefont{R.~R.~P.}
  \bibnamefont{Singh}}, \bibinfo{journal}{Phys. Rev. B}
  \textbf{\bibinfo{volume}{68}}, \bibinfo{pages}{094423}
  (\bibinfo{year}{2003}),
  \urlprefix\url{http://link.aps.org/doi/10.1103/PhysRevB.68.094423}.

\bibitem[{\citenamefont{Zheludev
  et~al.}(1998{\natexlab{c}})\citenamefont{Zheludev, Maslov, Tsukada,
  Zaliznyak, Regnault, Masuda, Uchinokura, Erwin, and
  Shirane}}]{Zheludev:98PRL}
\bibinfo{author}{\bibfnamefont{A.}~\bibnamefont{Zheludev}},
  \bibinfo{author}{\bibfnamefont{S.}~\bibnamefont{Maslov}},
  \bibinfo{author}{\bibfnamefont{I.}~\bibnamefont{Tsukada}},
  \bibinfo{author}{\bibfnamefont{I.}~\bibnamefont{Zaliznyak}},
  \bibinfo{author}{\bibfnamefont{L.~P.} \bibnamefont{Regnault}},
  \bibinfo{author}{\bibfnamefont{T.}~\bibnamefont{Masuda}},
  \bibinfo{author}{\bibfnamefont{K.}~\bibnamefont{Uchinokura}},
  \bibinfo{author}{\bibfnamefont{R.}~\bibnamefont{Erwin}}, \bibnamefont{and}
  \bibinfo{author}{\bibfnamefont{G.}~\bibnamefont{Shirane}},
  \bibinfo{journal}{Phys. Rev. Lett.} \textbf{\bibinfo{volume}{81}},
  \bibinfo{pages}{5410} (\bibinfo{year}{1998}{\natexlab{c}}).

\bibitem[{\citenamefont{Bogdanov and Shestakov}(1999)}]{Bogdanov:99}
\bibinfo{author}{\bibfnamefont{A.~N.} \bibnamefont{Bogdanov}} \bibnamefont{and}
  \bibinfo{author}{\bibfnamefont{A.~A.} \bibnamefont{Shestakov}},
  \bibinfo{journal}{Low Temperature Physics} \textbf{\bibinfo{volume}{25}},
  \bibinfo{pages}{76} (\bibinfo{year}{1999}).

\bibitem[{\citenamefont{Chovan et~al.}(2002)\citenamefont{Chovan, Papanicolaou,
  and Komineas}}]{Chovan:02}
\bibinfo{author}{\bibfnamefont{J.}~\bibnamefont{Chovan}},
  \bibinfo{author}{\bibfnamefont{N.}~\bibnamefont{Papanicolaou}},
  \bibnamefont{and} \bibinfo{author}{\bibfnamefont{S.}~\bibnamefont{Komineas}},
  \bibinfo{journal}{Phys. Rev. B} \textbf{\bibinfo{volume}{65}},
  \bibinfo{pages}{064433} (\bibinfo{year}{2002}),
  \urlprefix\url{http://link.aps.org/doi/10.1103/PhysRevB.65.064433}.

\bibitem[{\citenamefont{Bogdanov et~al.}(2004)\citenamefont{Bogdanov,
  R\"o\ss{}ler, Wolf, and M\"uller}}]{Bogdanov:04}
\bibinfo{author}{\bibfnamefont{A.~N.} \bibnamefont{Bogdanov}},
  \bibinfo{author}{\bibfnamefont{U.~K.} \bibnamefont{R\"o\ss{}ler}},
  \bibinfo{author}{\bibfnamefont{M.}~\bibnamefont{Wolf}}, \bibnamefont{and}
  \bibinfo{author}{\bibfnamefont{K.-H.} \bibnamefont{M\"uller}},
  \bibinfo{journal}{J. Magn. Mater.} \textbf{\bibinfo{volume}{272}},
  \bibinfo{pages}{332} (\bibinfo{year}{2004}).

\bibitem[{\citenamefont{Lumsden et~al.}(2001)\citenamefont{Lumsden, Sales,
  Mandrus, Nagler, and Thompson}}]{Lumsden2001}
\bibinfo{author}{\bibfnamefont{M.~D.} \bibnamefont{Lumsden}},
  \bibinfo{author}{\bibfnamefont{B.~C.} \bibnamefont{Sales}},
  \bibinfo{author}{\bibfnamefont{D.}~\bibnamefont{Mandrus}},
  \bibinfo{author}{\bibfnamefont{S.~E.} \bibnamefont{Nagler}},
  \bibnamefont{and} \bibinfo{author}{\bibfnamefont{J.~R.}
  \bibnamefont{Thompson}}, \bibinfo{journal}{Phys. Rev. Lett.}
  \textbf{\bibinfo{volume}{86}}, \bibinfo{pages}{159} (\bibinfo{year}{2001}).

\end{thebibliography}
\end{document}